\documentclass[
prb,twocolumn,floatfix,
amsmath,amssymb,superscriptaddress,bibnotes,longbibliography
]{revtex4-2}

\usepackage{hyperref}
\usepackage{graphicx}
\usepackage{mathtools}
\usepackage{color}
\usepackage{xcolor}
\usepackage{diagbox}
\usepackage{tabularx}
\newcolumntype{Y}{>{\centering\arraybackslash}X}

\usepackage{tabu}
\usepackage{verbatim}
\usepackage{amsmath}
\usepackage{stmaryrd}

\usepackage{tikz}

\usepackage[export]{adjustbox}
\usepackage{comment}
\usepackage{amssymb}

\newcommand{\T}{\mathcal{T}}
\newcommand{\ttau}{\urcorner}

\draft % marks overfull lines with a black rule on the right

\begin{document}
% --------------------------------------------------------------------

\title{Excitations and spectra from equilibrium real-time Green's functions}

\author{Xinyang Dong}
\affiliation{Department of Physics, University of Michigan, Ann Arbor, MI 48109, USA}
\author{Emanuel Gull}
\affiliation{Department of Physics, University of Michigan, Ann Arbor, MI 48109, USA}
\author{Hugo~U.~R.~Strand}
\affiliation{School of Science and Technology, Örebro University, SE-701 82 Örebro, Sweden}
\email{hugo.strand@oru.se}

\date{\today}

% --------------------------------------------------------------------
\begin{abstract}
The real-time contour formalism for Green's functions provides time-dependent information of quantum many-body systems.
In practice, the long-time simulation of systems with a wide range of energy scales is challenging due to both the storage requirements of the discretized Green's function
and the computational cost of solving the Dyson equation.
In this manuscript, we apply a real-time discretization based on a piece-wise high-order orthogonal-polynomial expansion to address these issues.
We present a superconvergent algorithm for solving the real-time equilibrium Dyson equation using the Legendre spectral method and the recursive algorithm for Legendre convolution.
We show that the compact high order discretization in combination with our Dyson solver enables long-time simulations using far fewer discretization points than needed in conventional multistep methods.
As a proof of concept, we compute the molecular spectral functions of H$_2$, LiH, He$_2$ and C$_6$H$_4$O$_2$ using self-consistent second-order perturbation theory and compare the results with standard quantum chemistry methods as well as the auxiliary second-order Green's function perturbation theory method.
\end{abstract}
% --------------------------------------------------------------------

\maketitle
\makeatletter
\let\toc@pre\relax
\let\toc@post\relax
\makeatother

% --------------------------------------------------------------------
\section{Introduction}
% --------------------------------------------------------------------

The finite-temperature real-time Green's function formalism of equilibrium quantum statistical mechanics \cite{Stefanucci:2013oq} has several advantages over the commonly used real-frequency and imaginary-time formalisms \cite{Mahan:2000kb}.
Unlike in the imaginary-time formalism, spectral functions can be extracted from real-time Green's functions without the need of ill-posed analytical continuation \cite{Jarrell:1996fj}.
Unlike in the real-frequency formalism, there is no explicit dependence on the location of poles on the real-axis. This allows one to solve self-consistent diagrammatic equations without further approximations \cite{Aulbur:2000aa, Martin:2016aa}, which may violate conservation laws \cite{Baym:1961ab, Baym:1962aa}.
However, to describe systems with disparate energy scales both high time resolution and long-time propagation are needed.

This requires a compact representation of data on the real-time axis in combination with an accurate solver for the real-time Green's function equation of motion, the Dyson equation.
The final time accessible, and hence the energy resolution of a simulation, critically depend on the efficiency with which this equation can be solved.
Current methods are built on an equidistant discretization in real-time which evolved from second order explicit methods \cite{KOHLER1999123, Stan:2009ab} to the current state-of-the-art $6^\text{th}$ order multistep method \cite{Schuler:2020aa}.
For two-time arguments, history truncation \cite{PhysRevB.105.115146} and matrix compression techniques have been developed \cite{Kaye:2021ab}, as well as adaptive time-stepping methods \cite{10.21468/SciPostPhysCore.5.2.030}.

In this paper we take a different approach and discretize the real-time axis in terms of a piece-wise high-order orthogonal-polynomial expansion on sequential panels.
Since the Green's function is smooth, the polynomial expansion converges exponentially with the expansion order $N$ \cite{boyd::2000}, yielding a compact representation.
Using the discretization to represent the mixed Green's function, we develop a superconvergent \cite{Wahlbin:1995} Dyson equation solver for equilibrium real-time propagation, with $2N^\text{th}$ order global convergence at the panel boundaries.
The compactness and high-order accuracy allows us to use large panels with comparably low polynomial degree and enables access to unprecedentedly long times.

As a proof-of-concept benchmark of the real-time panel representation and the high-order Dyson equation solver, we perform equilibrium real-time propagation using self-consistent second-order perturbation theory (GF2) \cite{PhysRevB.63.075112, Dahlen:2005aa, Phillips:2014aa, Phillips:2015ab, Kananenka:2016ut, Kananenka:2016ab, :/content/aip/journal/jcp/144/5/10.1063/1.4940900, :/content/aip/journal/jcp/145/20/10.1063/1.4967449, PhysRevB.100.085112} of molecules and compare our results to standard quantum chemistry methods and the recently developed approximation to GF2, Auxiliary GF2 (AGF2) \cite{Backhouse:2020aa, Backhouse:2020ab}. We also benchmark the state-of-the-art Nevanlinna analytical continuation method \cite{PhysRevLett.126.056402}.

This paper is organized as follows: In Sec.~\ref{sec:rtgf} we introduce the contour real-time Green's function formalism, the Dyson equation of motion, and their application to the case of equilibrium real-time propagation. The real-time panel discretization is introduced in Sec.~\ref{sec:discretization} with a preamble on imaginary-time discretization.
Using the compact representation, a high order algorithm for solving the Dyson equation of motion is developed in Sec.~\ref{sec:dyson_equation_solver}. The algorithmic asymptotic convergence properties and computational complexity are compared to state-of-the-art multistep methods in Sec.~\ref{sec:convergence} and Sec.~\ref{sec:complexity}. Proof-of-concept benchmarks on molecular systems using GF2 are shown in Sec.~\ref{sec:GF2}. Sec.~\ref{sec:conclusion} is devoted to conclusions and an outlook.

% --------------------------------------------------------------------

% --------------------------------------------------------------------
\section{Real-time Green's functions}
\label{sec:rtgf}
\begin{figure}
\includegraphics[scale=1] {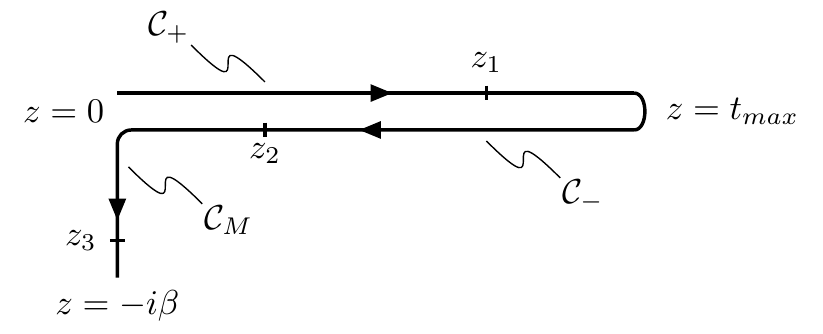}
\caption{Real-time contour for systems in initial thermal equilibrium at inverse temperature $\beta$.
\label{fig:contour}} \end{figure}

% --------------------------------------------------------------------
%
The general theory of non-equilibrium real-time Green's functions is built on the real-time contour formalism \cite{Stefanucci:2013oq}.
For systems that start in initial thermal equilibrium and evolve according to a time dependent Hamiltonian, the time propagation is performed along the L-shaped time contour $\mathcal{C}$, see Fig.\ \ref{fig:contour}.
The contour $\mathcal{C}$ consists of three branches $\mathcal{C} = \mathcal{C}_+ \cup \mathcal{C}_- \cup \mathcal{C}_M$, where the branch $\mathcal{C}_+$ is the forward propagation in real-time from the contour time $z=0$ to some maximal time $z = t_{\text{max}}$, $\mathcal{C}_-$ is the backward propagation in real-time, and $\mathcal{C}_M$ is the propagation in imaginary time to the final time $z = -i\beta$, where $\beta$ is the inverse temperature of the initial state \cite{Stefanucci:2013oq}.

% --------------------------------------------------------------------
% --------------------------------------------------------------------
%
In order to describe both thermal and temporal quantum correlations, we introduce the single particle Green's function $G$ that depends on two contour times $z$ and $z'$,
\begin{align}
    G_{ab}(z, z') = -i \langle T_{\mathcal{C}} c_a(z) c_b^\dagger(z') \rangle \, , \quad z, z' \in \mathcal{C} \, ,
\end{align}
where $T_{\mathcal{C}}$ is the contour time ordering operator, the operator $c_a^\dagger(z')$ ($c_b(z)$) creates (annihilates) an electron in the orbital $a$ ($b$) at the contour time $z'$ ($z$), and $\langle \cdot \rangle$ is an ensemble expectation value, see Ref.~\onlinecite{Stefanucci:2013oq}. In the following derivations we will suppress the orbital indices $a$ and $b$ for readability.

The equation of motion for the contour Green's function $G(z, z')$ is the integro-differential Dyson equation
\begin{align}
    (i S \partial_z - F(z))G(z, z') - \int_{\mathcal{C}} d\bar{z} \, \Sigma(z, \bar{z}) G(\bar{z}, z')
  = \delta_{\mathcal{C}}(z, z') \, ,
  \label{eq:contour_dyson}
\end{align}
where $S$ is the overlap matrix, $F(z)$ is the Fock matrix \cite{doi:10.1063/5.0003145}, $\Sigma(z, z')$ is the dynamic self-energy, and $\delta_\mathcal{C}$ is the contour Dirac-delta function \cite{Stefanucci:2013oq}.

By constraining the time arguments $z$ and $z'$ of the Green's function $G(z, z')$ to one of the three parts ($\mathcal{C}_+$, $\mathcal{C}_-$, $\mathcal{C}_M$) of the real-time contour $\mathcal{C}$, we can use the symmetry properties of $G$ to work with a reduced set of components.
One possible choice is
\begin{subequations}
\begin{align}
  &G^M(\tau - \tau') = -iG(z, z') \, , \quad z, z' \in \mathcal{C}_M ,
  \\
  &G^>(t, t') = G(z, z') \, , \quad z \in \mathcal{C}_- \, , \, \, z' \in \mathcal{C}_+ ,
  \label{eq:G_gtr_contour_component}
  \\
  &G^<(t, t') = G(z, z') \, , \quad z \in \mathcal{C}_+ \, , \, \, z' \in \mathcal{C}_- ,
  \label{eq:G_les_contour_component}
  \\
  &G^\ttau(t, \tau') = G(z, z') \, , \quad z \in \mathcal{C}_\pm \, , \, \, z' \in \mathcal{C}_M ,
\end{align}
\end{subequations}
where $G^M(\tau)$ is the imaginary time, $G^\gtrless(t, t')$ the greater/lesser, and $G^\ttau(t, \tau')$ the mixed Green's function.
For the resulting coupled Dyson equations for this set of Green's function components see Ref.~\onlinecite{Aoki:2014kx}.
%

% --------------------------------------------------------------------
\subsection{Equilibrium real-time Green's functions}
\label{sec:eqrtgf}
% --------------------------------------------------------------------
%
We will only consider the case of equilibrium real-time evolution \cite{Strand:2015ac, Kaye:2021aa}, when the time evolution of the system is governed by the same time-independent Hamiltonian as the initial thermal equilibrium state. In this case the greater and lesser Green's functions $G^\gtrless$ are time translation invariant, $G^\gtrless(t, t') = G^\gtrless(t - t')$, and can be inferred from the mixing Green's function $G^\ttau$ as
\begin{equation}
  G^<(t) = G^\ttau(t, 0)
  \, , \quad
  G^>(t) = \xi G^\ttau(t, \beta)
  \, ,
  \label{eq:lessergreater_from_mixing}
\end{equation}
where $\xi = -1 (+1)$ for fermions (bosons).
In equilibrium, the spectral function $A(\omega)$, related to the photo-emission spectrum, is given by the retarded Green's function $G^R$ in real-frequency,
\begin{equation}
A(\omega) = -\frac{1}{\pi} \text{Im} \left[ G^R(\omega) \right]
\, .
\label{eq:A_omega_from_GR}
\end{equation}
In real-time $G^R$ is determined by $G^\gtrless$ through the relation
\begin{align}
G^R(t)
&= \theta(t)\left( G^>(t) - G^<(t) \right) \nonumber \\
&= - \theta(t)\left( G^\ttau(t, \beta) + G^\ttau(t, 0) \right)
\, .
\label{eq:GR_from_Gtv}
\end{align}
Therefore, with an initial state determined by $G^M$, all real time behavior can be determined by $G^\ttau(t, \tau')$, and it is sufficient to solve the Dyson equations for $G^M$ and $G^\ttau$.

In imaginary time the Dyson equation for $G^M(\tau)$ is given by \cite{Aoki:2014kx}
\begin{align}
(-S\partial_{\tau} - F) G^M(\tau)
- \int_0^{\beta} d\bar{\tau} \, \Sigma^M(\tau - \bar{\tau}) G^M(\bar{\tau}) = 0 \, ,
\label{eq:mat_dyson}
\end{align}
with the boundary condition $[G^M(0) + G^M(\beta)]S = -\mathbf{1}$ and the dynamic imaginary-time self-energy $\Sigma^M(\tau)$.
For the connection to the imaginary-frequency Matsubara formalism see Ref.~\onlinecite{doi:10.1063/5.0003145}.
The solution $G^M(\tau)$ provides the initial condition
\begin{equation}
  G^\ttau(0, \tau) = -i G^M(\beta - \tau)
%  \, ,
  \label{eq:mix_initial_cond}
\end{equation}
for the real-time evolution of $G^\ttau$.
The Dyson equation for $G^\ttau$ is given by
\begin{equation}
(i S \partial_t - F) G^\ttau(t, \tau) - \int_0^t d\bar{t} \, \Sigma^R(t - \bar{t}) G^\ttau(\bar{t}, \tau) = Q^\ttau(t, \tau)
  \label{eq:mix_dyson}
  \, ,
\end{equation}
where the right hand side $Q^\ttau$ accounts for the temporal correlations with the initial imaginary time state
\begin{align}
  Q^\ttau(t, \tau) = \int_0^\beta d\bar{\tau} \, \Sigma^\ttau(t, \bar{\tau}) G^M(\bar{\tau} - \tau)
%  \, ,
  \label{eq:mix_dyson_rhs}
\end{align}
through its dependence on $G^M(\tau)$. In Eq.\ (\ref{eq:mix_dyson}) the retarded self-energy integral kernel $\Sigma^R$ is given by
\begin{equation}
  \Sigma^R(t) = \xi \Sigma^\ttau(t, \beta) - \Sigma^\ttau(t, 0) \, , \quad t \ge 0
  \, .
\label{eq:sigma_r}
\end{equation}

Combining the Dyson equations (\ref{eq:mat_dyson}) and (\ref{eq:mix_dyson}) with expressions for the self energies
\begin{equation}
  \Sigma^M = \Sigma^M[G^M]
  \, , \quad
  \Sigma^\ttau = \Sigma^\ttau[G^\ttau]
  \label{eq:sigma_func}
\end{equation}
gives the closed set of equations (\ref{eq:mat_dyson}), (\ref{eq:mix_dyson}) and (\ref{eq:sigma_func}) that can be solved first in imaginary time for $G^M(\tau)$ and then in real time for $G^\ttau(t, \tau)$.
For an explicit example of such a self-energy expression see Eq.~(\ref{eq:mix_sigma_gf2}) in Sec.~\ref{sec:GF2} where the self-consistent second-order self-energy approximation (GF2)
\cite{PhysRevB.63.075112, Dahlen:2005aa, Phillips:2014aa, Phillips:2015ab, Kananenka:2016ut, Kananenka:2016ab, :/content/aip/journal/jcp/144/5/10.1063/1.4940900, :/content/aip/journal/jcp/145/20/10.1063/1.4967449, PhysRevB.100.085112} is introduced.

% --------------------------------------------------------------------
\section{High-order discretization}
\label{sec:discretization}
% --------------------------------------------------------------------

Solving the Dyson equations [Eqs.~\ref{eq:mat_dyson} and \ref{eq:mix_dyson}] numerically requires a precise representation of the imaginary time Green's function $G^M(\tau)$ and self-energy $\Sigma^M(\tau)$ as well as their mixed counter parts $G^\ttau(t, \tau)$, $\Sigma^\ttau(t, \tau)$.

For entire functions, finite orthogonal polynomial expansions converge supergeometrically \cite{boyd::2000}, i.e.\ the polynomial coefficients decay faster than exponentially with polynomial order. Stable high-order integro-differential solvers can be formulated for this class of expansions \cite{Jie-Shen:2011uq, Olver:2020vo}.
Building on our previous work \cite{doi:10.1063/5.0003145}, we use a high order Legendre expansion for the imaginary time $\tau$-axis. The real-time $t$-axis is subdivided into panels, each containing a Legendre expansion in real-time.
This panel approach is inspired by the higher order elements employed in the spectral/hp element methods in computational fluid dynamics \cite{Pozrikidis:2014aa,Karniadakis:1999aa}.
For the mixed Green's function $G^\ttau(t, \tau)$ we use a direct product basis of the imaginary-time and real-time representations.

We employ the Legendre polynomial basis since it is possible to express the Fredholm and Volterra integrals in Eqs.~\ref{eq:mat_dyson} and \ref{eq:mix_dyson} directly in Legendre coefficient space, using a recursive algorithm \cite{Hale:2014ab}.

% --------------------------------------------------------------------
\subsection{Imaginary-time Legendre polynomial expansion}
% --------------------------------------------------------------------

We represent the functions $G^M(\tau)$ and $\Sigma^M(\tau)$ with one imaginary time argument $\tau$ using a finite Legendre polynomial expansion of order $N_\tau$
\begin{align}
  G^M(\tau) \approx \sum_{m=0}^{N_\tau-1} G^M_m P_m[\psi_M(\tau)] \, ,
  \label{eq:taubasis}
\end{align}
where $G^M_{m}$ are the Legendre polynomial expansion coefficients of $G^M(\tau)$, $P_m(x)$ is the Legendre polynomial of order $m$ defined on $x \in [-1,1]$, and the linear function
\begin{equation}
  \psi_M(\tau) = \frac{2\tau}{\beta} - 1
  \label{eq:tau_psi}
\end{equation}
maps imaginary time $\tau \in [0, \beta]$ to $\psi_M(\tau) \in [-1, 1]$.
For Green's functions and self-energies the expansions converge faster than exponential with $N_\tau$ since both these classes of functions are infinitely derivable in imaginary time $\tau$ \cite{Boehnke:2011fk, doi:10.1063/5.0003145}.

We note that several other representations for imaginary time have been explored, including power mesh discretizations \cite{PhysRevLett.89.126401, KuPhD2000, Stan:2006vc, Stan:2009aa, PhysRevB.97.115164}, pole expansions \cite{PhysRevLett.110.146403, PhysRevB.88.075105}, spline grids \cite{Kananenka:2016ab}, Chebyshev orthogonal polynomials \cite{PhysRevB.98.075127}, numerical basis functions from singular value decomposition of the analytical continuation kernel (also known as the intermediate representation basis) \cite{PhysRevB.96.035147, PhysRevB.98.035104, CHIKANO2019181}, and analytical basis functions from interpolative decomposition of the same kernel \cite{2021arXiv210713094K, 2021arXiv211006765K}.

% --------------------------------------------------------------------
\subsection{Imaginary time Dyson equation solver}
% --------------------------------------------------------------------

The solution of the imaginary-time Dyson equation (\ref{eq:mat_dyson}) can be formulated directly in terms of the Legendre polynomial coefficients $G_m^M$ in Eq.\ \ref{eq:tau_psi}, as shown in Ref.~\onlinecite{doi:10.1063/5.0003145}.

The resulting linear system can be solved iteratively for $G_m^M$ using the generalized minimal residual algorithm (GMRES) \cite{doi:10.1137/0907058} and a Matsubara-frequency sparse sampling \cite{PhysRevB.101.035144, PhysRevB.101.205145} preconditioner using the Legendre sparse sampling points derived in Appendix~\ref{app:SparseSampling}.
The imaginary time convolution integral in Eq.~(\ref{eq:mat_dyson}) is computed in the Legendre coefficient space using the recursive convolution method \cite{Hale:2014ab, doi:10.1063/5.0003145}, scaling as $\mathcal{O}(N^2)$ with the polynomial order $N$.
The recursive convolution method, in combination with the preconditioned iterative linear solver, gives a Dyson equation solver algorithm with the same $\mathcal{O}(N^2)$ quadratic scaling.

For the benchmarks presented in Sec.\ \ref{sec:GF2} the solution from the Matsubara sparse sampling method also reaches machine precision when using roughly twice the number of Legendre coefficients required by the Legendre spectral algorithm of Ref.\ \onlinecite{doi:10.1063/5.0003145}.

% --------------------------------------------------------------------
\subsection{Real-time Legendre panel expansion}
\label{sec:panel_exp}
% --------------------------------------------------------------------

\begin{figure}[h!]
\ \\ \ \\[-1mm]
  \includegraphics[scale=1] {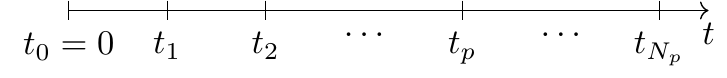} %\\[-3mm]
\caption{Real-time panel representation. \label{fig:panel}
} \end{figure}

To represent functions with one real time argument, like the retarded self-energy $\Sigma^R(t)$, we construct panels by dividing the real-time $t$-axis using equidistant points $t_p = \Delta t \cdot p$, with $p = 0, 1, ..., N_p$, see Fig.\ \ref{fig:panel}.

For this segmentation we define the real-time panels $\T_p$ as the sub intervals
\begin{align*}
    \T_p \equiv [t_p, t_{p+1}]
    = [p \Delta t, (p+1) \Delta t]
    \, , \quad p \in 0, 1, ..., N_p - 1
    \, .
\end{align*}
For $t$ on a given panel $p$, $t \in \T_p$, the real-time dependent self-energy $\Sigma^R(t)$ can be discretized using the finite Legendre expansion $\Sigma^R_p(t)$ on panel $p$
\begin{equation}
    \Sigma^R(t) \approx
    \Sigma^R_p(t)
    =
    \sum_{n=0}^{N_t-1} \Sigma^R_{p,n} P_n[\psi_p(t)]
    \, ,
    %\quad \text{for } t \in \T_p \, ,
    \label{eq:sigma_R_panel}
\end{equation}
where $\Sigma^R_{p,m}$ are the Legendre coefficients, and the linear function
%
%$\psi_p(t) = 2\left(t/(\Delta t) - p\right) - 1$,
%
\begin{align}
    \psi_p(t) = 2\left(\frac{t}{\Delta t} - p\right) - 1
    \label{eq:t_psi}
\end{align}
maps times $t \in \T_p$ back to the interval $\psi_p(t) \in [-1, 1]$.

The self-energy $\Sigma^R(t)$ for all $t$ can be expressed as the direct sum of the panel expansions $\Sigma^R_p(t)$
\begin{align}
    \Sigma^R(t) \approx
    \sum_{p=0}^{N_p - 1}
    \Sigma^R_p(t)
    =
    \sum_{p=0}^{N_p - 1}
    \left[ \,
    \sum_{n=0}^{N_t-1} \Sigma^R_{p,n} P_n[\psi_p(t)]
    \right] \, ,
    \label{eq:tbasis}
\end{align}
by defining $P_n(x) = 0$ for $x \notin [-1, 1]$.

% --------------------------------------------------------------------
\subsection{Imaginary- and real-time product basis}
% --------------------------------------------------------------------

\begin{figure}
\includegraphics[scale=1] {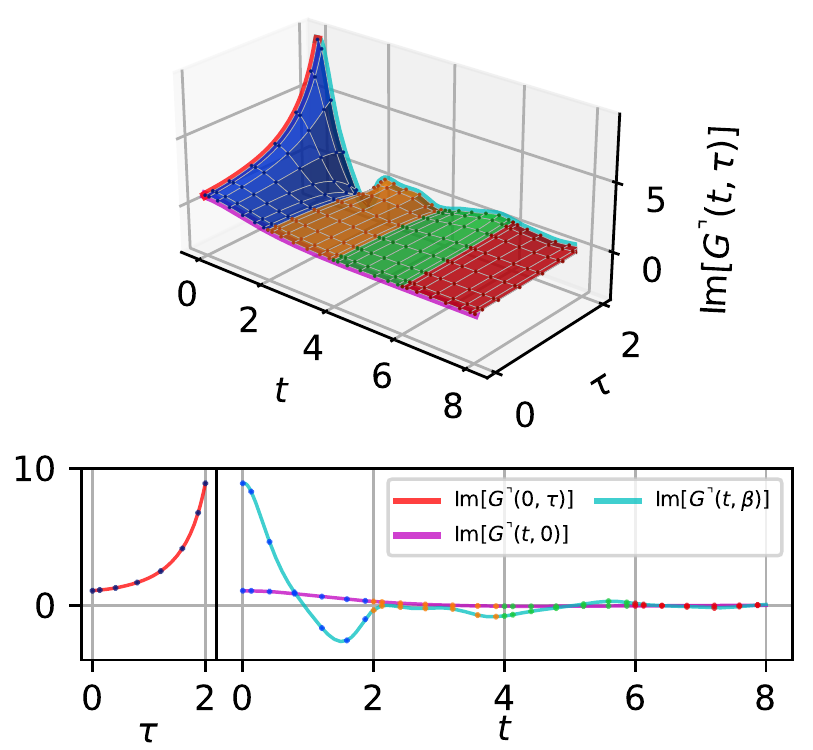}
\caption{
  Real- and imaginary-time panel representation of the mixed Green's function $G^\ttau(t, \tau)$
  for a H$_2$ dimer
  at hydrogen distance $r=0.5\,$Å
  and (for visualization purposes)
  the low inverse temperature
  $\beta = 2\,$Ha$^{-1}$.
  Upper panel:
  Product representation using four panels (blue, yellow, green, and red surfaces). Each panel has a product basis with eight order polynomials both in time $t$ and imaginary-time $\tau$, $N_t=8$, $N_\tau=8$. The corresponding collocation nodes are also shown (dark circle markers).
  Lower left panel:
  Initial imaginary time solution at $t=0$, where Im$[G^\ttau(0, \tau)] = - G^M(\beta - \tau)$ in terms of its polynomial expansion (red) and collocation nodes (dark blue).
  Lower right panel:
  Time evolution of $G\ttau(t, \tau)$ at $\tau = 0$ and $\beta$ (cyan and magenta lines respectively) related to the lesser and greater Green's functions, see Eq.~\ref{eq:lessergreater_from_mixing}.
  \label{fig:panel_repr}
} \end{figure}

Since the equilibrium real-time evolution is described by the mixed Green's function $G^\ttau(t, \tau)$, we combine the Legendre expansion in imaginary-time $\tau$ and the Legendre panel based expansion in real-time $t$ by forming a direct product basis of Eq.~\ref{eq:taubasis} and \ref{eq:tbasis}. The resulting representation of $G^\ttau(t, \tau)$ takes the form
\begin{multline}
    G^\ttau(t, \tau) \approx
    \sum_{p=0}^{N_p-1}
    G^\ttau_p(t, \tau)
    \\ =
    \sum_{p=0}^{N_p-1}
    \left[ \,
    \sum_{n=0}^{N_t-1} \sum_{m=0}^{N_\tau-1} G^\ttau_{p,nm} P_n[\psi_p(t)] P_m[\psi_M(\tau)]
    \right] \, ,
    \label{eq:gttau_product_basis}
\end{multline}
where $G^\ttau_{p,nm}$ is a rank-3 tensor of Legendre polynomial coefficients.
Since $G^\ttau(t, \tau)$ is an entire function, the polynomial coefficients decay supergeometrically \cite{boyd::2000} with $n$ and $m$, and the discretization converges faster than exponentially with respect to $N_t$ and $N_\tau$.
As an example the product representation of $G^\ttau(t, \tau)$ for the hydrogen dimer is shown in Fig.~\ref{fig:panel_repr}.

% --------------------------------------------------------------------
\subsection{Legendre collocation points}
% --------------------------------------------------------------------
%
While we will solve the Dyson equation in Legendre coefficient space, it is also important to be able to transform between the Legendre coefficients of the Green's function $G^\ttau_{p,nm}$ and the Green's function $G^\ttau_{p}(t_{p, i}, \tau_j)$ on a grid of real-time $t_{p, i}$ and imaginary-time $\tau_j$, for evaluating self-energy with approximations given by direct products of Green's functions in time, like GF2.
For this purpose we use a set of \emph{collocation points} \cite{Jie-Shen:2011uq} that have stable linear transformations from and to Legendre coefficient space.

In this work we use the Legendre-Gauss-Lobatto collocation points \cite{Jie-Shen:2011uq} $x^{(N)}_i$ given by the roots of $(1-x^2)P_N(x) = 0$ for $0 < i < N-1$  and the points at the interval boundaries $x^{(N)}_0 = -1$, $x^{(N)}_N = 1$.
The linear transforms are given by the Legendre Vandermonde matrix $P^{(N)}_{in}$ and its inverse $S^{(N)}_{ni}$
\begin{equation}
  P^{(N)}_{in} = P_n(x^{(N)}_i)
  \, , \quad
  S^{(N)}_{ni} = \frac{\omega_i}{W_n} P_n(x^{(N)}_i)
  \, ,
\end{equation}
where $W_n = 2/(2n + 1) = \int_{-1}^1 P_n^2(x) \, dx$ and $\omega_i = 2/(N^2+N) \cdot P_N^{-2}(x_i)$ \cite{Jie-Shen:2011uq}.

Given the collocation points on the fundamental interval $[-1, 1]$, the real- and imaginary-time collocation points $t_{p, i}$ and $\tau_j$ are given by the inverse of the linear maps in Eqs.~(\ref{eq:tau_psi}) and (\ref{eq:t_psi})
\begin{equation}
  t_{p, i} = \psi^{-1}_p(x^{(N_t)}_i)
  \, , \quad
  \tau_j = \psi^{-1}_M(x^{(N_\tau)}_j)
  \, .
\end{equation}
The explicit linear transformations take the form
\begin{equation}
  G^M(\tau_j) = \sum_{m=0}^{N_\tau-1} P^{(N_\tau)}_{jm} G^M_m
  \, , \quad
  G^M_m = \sum_{j=0}^{N_\tau-1} S^{(N_\tau)}_{mj} G^M(\tau_j)
\end{equation}
for the imaginary time Green's function $G^M$, and for the mixed Green's function $G^\ttau$ the product basis gives
\begin{align}
  &G^\ttau(t_{p,i}, \tau_j) = \sum_{n=0}^{N_t-1} \sum_{m=0}^{N_\tau-1} P^{(N_t)}_{in} P^{(N_\tau)}_{jm} G^\ttau_{p, nm}
  \, , \nonumber \\
  &G^\ttau_{p, nm} = \sum_{i=0}^{N_t-1} \sum_{j=0}^{N_\tau-1} S^{(N_t)}_{ni} S^{(N_\tau)}_{mj} G^\ttau(t_{p, i}, \tau_j)
  \, ,
\end{align}
together with analogous relations for the self-energy components $\Sigma^M$ and $\Sigma^\ttau$.

% --------------------------------------------------------------------
\section{Dyson equation solver}
\label{sec:dyson_equation_solver}
% --------------------------------------------------------------------

Given the real-time panel discretization [Fig.\ \ref{fig:panel}] and the Legendre real- and imaginary-time product basis [Eq.\ \ref{eq:gttau_product_basis}], we will now reformulate the Dyson equation [Eq.\ \ref{eq:mix_dyson}] in Legendre coefficient space, also known as a Legendre \emph{spectral} formulation \cite{Jie-Shen:2011uq}.
In section \ref{sec:history_integral} we first express the history integral term in the Dyson equation (\ref{eq:mix_dyson}) using the real-time panel representation of $\Sigma^R$ and $G^\ttau$.
Then, in section \ref{sec:fast_conv}, we adapt the recursive Legendre convolution algorithm \cite{Hale:2014ab} to evaluate each non-zero combination of self-energy and Green's function panels in the integral.
Finally, in section \ref{sec:legendre_dyson} we map the remaining terms in the Dyson equation (\ref{eq:mix_dyson}) to Legendre coefficient space, arriving at a complete Legendre spectral formulation.

% --------------------------------------------------------------------
\subsection{Real-time panel history integral}
\label{sec:history_integral}
% --------------------------------------------------------------------

Using the real-time panel notations in section \ref{sec:panel_exp}, the history integral $I(t, \tau)$ in the Dyson equation (\ref{eq:mix_dyson})
\begin{equation}
  I(t, \tau)
  =
  \int_0^t d\bar{t} \, \Sigma^R(t - \bar{t}) G^\ttau(\bar{t}, \tau)
%  \, ,
\end{equation}
can be written as a sum of functions $I_p(t, \tau)$ supported on panel $p$, i.e.\
\begin{equation}
  I(t, \tau)
  =
  \sum_{p=0}^{N_p-1} I_p(t, \tau)
  \, , \quad
  \text{with } I_p(t, \tau) \equiv 0 \, , \, \forall t \notin \T_p
  \, .
  \label{eq:Ip_sum}
\end{equation}
Each function $I_p(t, \tau)$ can in turn be written as a sum of integrals over the mixed Green's function's panel components $G^\ttau_p(t, \tau)$ defined in Eq.~(\ref{eq:gttau_product_basis})
\begin{equation}
  I_p(t, \tau)
  =
  \sum_{k=0}^{p}
  \int_0^t
  d\bar{t} \, \Sigma^R(t - \bar{t}) G_k^\ttau(\bar{t}, \tau)
%  \equiv
%  \sum_{k=0}^{p}
%  I_{p, k}(t, \tau)
  \, ,
  \label{eq:Ip}
\end{equation}
where $G^\ttau_k(\bar{t}, \tau)$ is non-zero for $\bar{t} \in \T_k$.
The finite support in $t$ and $\bar{t}$ restricts the integration argument of $\Sigma^R(t - \bar{t})$ in each term of Eq.\ (\ref{eq:Ip}) since
\begin{equation}
  t \in \T_p
  \text{ and }
  \bar{t} \in \T_k
  \quad \Rightarrow \quad
  t - \bar{t} \in \T_{p-k-1} \bigcup \T_{p-k}
  \, .
\end{equation}
Hence, only two $\Sigma^R$-panels contribute in Eq.\ (\ref{eq:Ip})
\begin{align}
  &t \in \T_p \Rightarrow \nonumber \\
  &\int_0^t
  d\bar{t} \, \Sigma^R(t - \bar{t}) G_k^\ttau(\bar{t}, \tau) \nonumber
  \\ &=
  \int_0^t \!\!\! d\bar{t} \, \Sigma^R_{p-k-1}(t \! - \! \bar{t}) G_k^\ttau(\bar{t}, \tau)
  + \!\!
  \int_0^t \!\!\! d\bar{t} \, \Sigma^R_{p-k}(t \! - \! \bar{t}) G_k^\ttau(\bar{t}, \tau) \nonumber
  \\ &=
%  I^>_{p-k-1, k}(t, \tau) + I^<_{p-k, k}(t, \tau)
%  I^>_{p-k-1, k}(t, \tau) + I^<_{p-k, k}(t, \tau)
  %
  \Sigma^R_{p-k-1} \overset{>}{\ast} G_k^\ttau
  +
  \Sigma^R_{p-k} \overset{<}{\ast} G_k^\ttau
  \, ,
\end{align}
where in the last step we have introduced a short notation to represent these two types of panel integrals.
Using this short hand notation, the history integral $I_p(t, \tau)$ in Eq.\ (\ref{eq:Ip}) can be written as
\begin{equation}
  I_p(t, \tau)
  =
  \mathcal{I}_p(t, \tau) + \Sigma^R_p \overset{<}{\ast} G^\ttau_0 + \Sigma^R_0 \overset{<}{\ast} G^\ttau_p
  \, ,
  \label{eq:panel_hist}
\end{equation}
where the integrals depending on $\Sigma^R_p$ and $G^\ttau_p$ have been separated from the integrals over earlier panels
\begin{equation}
  \mathcal{I}_p(t, \tau) \equiv
  \sum_{k=0}^{p-1} \left( \Sigma^R_{p-1-k} \overset{>}{\ast} G^\ttau_k \right)
  +
  \sum_{k=1}^{p-1} \left( \Sigma^R_{p-k} \overset{<}{\ast} G^\ttau_k \right)
  \, .
  \label{eq:hist_int}
\end{equation}
The separation in Eq.\ (\ref{eq:panel_hist}) is prepared for direct use in the panel formulation of Dyson equation (\ref{eq:mix_dyson}), where $G^\ttau_p$ will be solved for and $\Sigma^R_p$ will be iteratively updated using Eq.\ (\ref{eq:sigma_func}).

% --------------------------------------------------------------------
\subsection{Legendre-spectral panel integrals}
\label{sec:fast_conv}
% --------------------------------------------------------------------

For the panel-history integral $I_p(t, \tau)$, the two types of panel integrals appearing in Eq.\ (\ref{eq:Ip}) can be readily computed in Legendre coefficient space.

In both cases the integral bounds are determined by the support of
the $\Sigma^R$ and $G^\ttau$ panel components
\begin{align}
  \Sigma^R_{p-k-1} &\overset{>}{\ast} G_k^\ttau
  \equiv
  \int_0^t d\bar{t} \, \Sigma^R_{p-k-1}(t - \bar{t}) G_k^\ttau(\bar{t}, \tau) \nonumber
  \\
  =
  &\int_{\max (t_k, t - t_{p-k})}^{\min (t_{k+1}, t - t_{p-k-1})} d\bar{t} \,
  \Sigma^R_{p-k-1}(t - \bar{t}) G_k^\ttau(\bar{t}, \tau) \nonumber
  \\
  =
  &\int_{t - t_{p-k}}^{t_{k+1}} d\bar{t} \, \Sigma^R_{p-k-1}(t - \bar{t}) G_k^\ttau(\bar{t}, \tau)
  \, ,
  \label{eq:panel_gtr}
\end{align}
and analogously
\begin{align}
  \Sigma^R_{p-k} \overset{<}{\ast} G_k^\ttau
  &\equiv
  \int_0^t d\bar{t} \, \Sigma^R_{p-k}(t - \bar{t}) G_k^\ttau(\bar{t}, \tau) \nonumber
  \\
  &=
  \int_{t_k}^{t - t_{p-k}} d\bar{t} \, \Sigma^R_{p-k}(t - \bar{t}) G_k^\ttau(\bar{t}, \tau)
  \, .
  \label{eq:panel_les}
\end{align}
Hale and Townsend \cite{Hale:2014ab} have derived a recursive method for this kind of Volterra type convolution integrals, with an external time argument in the integration bounds and in the integration kernel $\Sigma^R$.

The linear operator corresponding to the integration and the integration kernel is given by
\begin{equation}
  [\Sigma^R_q \overset{\lessgtr}{\ast}] = \frac{\Delta t}{2} B^\lessgtr[\Sigma^R_q]
  \, ,
  \label{eq:convol_B}
\end{equation}
where $B^\lessgtr$ is a matrix in Legendre coefficient space constructed via the recursion relation \cite{doi:10.1063/5.0003145}
\begin{equation}
  B^\lessgtr_{n, m+1} =
  - \frac{2m + 1}{2n + 3} B^\lessgtr_{n+1, m}
  + \frac{2m + 1}{2n - 1} B^\lessgtr_{n-1, m}
  + B^\lessgtr_{n, m-1}
%  \, .
  \label{eq:convol_recur}
\end{equation}
and the starting relations
\begin{align}
  &B^\lessgtr_{n, 0} =
    \left\{
  \begin{array}{lr}
    f_0 \mp \frac{f_1}{3} \, , & n = 0 \\[2mm]
    \pm (\frac{f_{n-1}}{2n - 1} - \frac{f_{n+1}}{2n + 3}) \, , & \quad n \geq 1
  \end{array}
  \right. \nonumber
  \\
  &B^\lessgtr_{n, 1} =
  \mp B^\lessgtr_{n, 0} + \frac{B^\lessgtr_{n-1, 0}}{2n - 1} - \frac{B^\lessgtr_{n+1, 0}}{2n + 3}
  \, , \quad n \ge 1
  \label{eq:starting}
\end{align}
with the special case, $B^\lessgtr_{0, 1} = - B^\lessgtr_{1, 0} / 3$, for $n = 0$.
The recursion relation in Eq.\ (\ref{eq:convol_recur}) is only stable in the lower triangular part of the coefficient matrix, and the upper triangular coefficients are computed from
\begin{align}
    B^\lessgtr_{n,m} = (-1)^{m+n} \frac{2n + 1}{2m + 1} B^\lessgtr_{m,n}
    \, .
    \label{eq:convol_trans}
\end{align}
In the case of Eq.\ (\ref{eq:convol_B}), the Legendre coefficient vector $f_n$ in Eq.\ (\ref{eq:starting}) is given by the real-time panel Legendre coefficients of the self energy $\Sigma^R$ on panel $q$ [Eq.\ (\ref{eq:sigma_R_panel})], $f_n = \Sigma^R_{q,n}$.

Using the integral operator construction of Eq.\ (\ref{eq:convol_B}), the product basis Legendre coefficients $I_{p, nm}$ [Eq.\ \ref{eq:gttau_product_basis}] of the history integral $I_p(t, \tau)$ can be calculated using matrix products in Legendre coefficient space according to
\begin{align}
  I_{p, nm} =
  \mathcal{I}_{p,nm}
  &+
  \sum_{n'=0}^{N_t-1}
  %\left(
  [\Sigma^R_{p} \overset{<}{\ast}]_{nn'} G^\ttau_{0, n'm} \nonumber
  \\ &+
   \sum_{n'=0}^{N_t-1}
   [\Sigma^R_{0} \overset{<}{\ast}]_{nn'} G^\ttau_{p, n'm}
   %  \right)
  \, ,
  \label{eq:I_hat_coeff}
\end{align}
where $G^\ttau_{p, nm}$ are the real-time panel Legendre coefficients of $G^\ttau$ on panel $p$, and $\mathcal{I}_{p,nm}$ are the product basis Legendre coefficients of $\mathcal{I}_p(t, \tau)$ in Eq.\ (\ref{eq:hist_int}) given by
\begin{align}
  \mathcal{I}_{p,nm}
  &=
  \sum_{k=0}^{p-1}
  \sum_{n'=0}^{N_t-1}
  [\Sigma^R_{p-1-k} \overset{>}{\ast}]_{nn'} G^\ttau_{k,n'm}
  \nonumber \\
  &+
  \sum_{k=1}^{p-1}
  \sum_{n'=0}^{N_t-1}
  [\Sigma^R_{p-k} \overset{<}{\ast}]_{nn'} G^\ttau_{k,n'm}
  \, .
  \label{eq:mix_dyson_panel_legendre_history}
\end{align}
%

% --------------------------------------------------------------------
\subsection{Real-time panel right hand side}
% --------------------------------------------------------------------

To formulate the real-time Dyson equation (\ref{eq:mix_dyson}) using the real-time panel representation,
the right hand side $Q^\ttau(t,\tau)$ in equation [Eq.~(\ref{eq:mix_dyson})] also has to be expressed as a sum of panel restricted functions
\begin{equation}
  Q^\ttau(t, \tau) = \sum_{p=0}^{N_p-1} Q^\ttau_p(t, \tau)
  \, , \quad
  \text{with } Q^\ttau_p(t, \tau) \equiv 0 \, , \, \forall t \notin \T_p
  \, ,
  \label{eq:Qp_sum}
\end{equation}
where $Q^\ttau_p(t, \tau)$ is given by
\begin{equation}
  Q_p^\ttau(t, \tau)
  =
  \int_0^\beta d\bar{\tau} \, \Sigma^\ttau_p(t, \bar{\tau}) G^M(\bar{\tau} - \tau)
  \, .
  \label{eq:Qp}
\end{equation}
This class of integrals can be computed in imaginary time Legendre coefficient space using the recursive algorithm of Eq.\ (\ref{eq:convol_recur}) as shown in Ref.\ \onlinecite{doi:10.1063/5.0003145}.
Accounting for the sign in the convolution argument of Eq.\ (\ref{eq:Qp}) gives the real-time panel Legendre coefficients of $Q^\ttau_p$ as
\begin{align}
    Q^\ttau_{p,nm}
    =
    %(-1)^{m+1} \sum_{m'}  [G^M{\ast}]_{mm'} \Sigma^\ttau_{p,nm'}
    \sum_{m'}  [\tilde{G}^M{\ast}]_{mm'} \Sigma^\ttau_{p,nm'}
    \, ,
    \label{eq:Qp_leg}
\end{align}
where the integral operator $[\tilde{G}^M {\ast}]$ is given by
\begin{equation}
  [\tilde{G}^M {\ast}] = \frac{\beta}{2} \left( B^<[\tilde{G}^M] + \xi B^>[\tilde{G}^M] \right)
  \label{eq:mat_B}
\end{equation}
with $B^\lessgtr$ given by Eqs.\ (\ref{eq:convol_recur}), (\ref{eq:convol_trans}) and  (\ref{eq:starting}) using the modified Legendre coefficients $f_n = \tilde{G}^M_n = (-1)^{n+1} G^M_n$.
See Appendix\ \ref{app:imtime} for a derivation.

% --------------------------------------------------------------------
\subsection{Legendre-spectral panel Dyson equation}
\label{sec:legendre_dyson}
% --------------------------------------------------------------------

With all the terms appearing in the real-time Dyson equation (\ref{eq:mix_dyson}) expressed on the panel subdivision of the real-time axis, we are now in a position to formulate the corresponding real-time \emph{panel} Dyson equation for
the mixing Green's function panel component $G^\ttau_p(t, \tau)$, with $t \in \T_p$.

Using the panel expression for both the history integral $I(t, \tau)$ in Eq.~(\ref{eq:Ip_sum}) and Eq.\ (\ref{eq:panel_hist}), and the right-hand side $Q^\ttau(t, \tau)$ in Eq.\ (\ref{eq:Qp_sum}), the real-time panel Dyson equation for $G^\ttau_p(t, \tau)$ becomes
\begin{multline}
  (i S \partial_t - F - \Sigma^R_0 \overset{<}\ast ) G^\ttau_p(t, \tau)
  \\ =
  Q^\ttau_p(t, \tau) + \mathcal{I}_p(t, \tau) + [\Sigma^R_p \overset{<}\ast G^\ttau_0](t, \tau)
  \, ,
  \label{eq:mix_dyson_panel}
\end{multline}
with the boundary conditions
\begin{align}
  G^\ttau_p(0, \tau) &= i\xi G^M(\beta - \tau) \, , &\text{for } p = 0 \, ,
  \nonumber \\
  G^\ttau_p(t_p, \tau) &= G^\ttau_{p-1}(t_p, \tau) \, , &\text{for } p > 0 \, ,
  \label{eq:mix_dyson_panel_bc}
\end{align}
given by the initial boundary condition in Eq.\ (\ref{eq:mix_initial_cond}) and the continuity of $G^\ttau(t, \tau)$ between panels.

We now reformulate all terms in Eqs.~(\ref{eq:mix_dyson_panel}) and (\ref{eq:mix_dyson_panel_bc}) in the panel Legendre product basis. The goal is to translate all expressions containing $G^\ttau_p(t, \tau)$ in terms of the polynomial coefficients $G^\ttau_{p,nm}$ defined in Eq.~(\ref{eq:gttau_product_basis}).

%Using $P_n(\pm 1) = (\pm 1)^n$,
%the boundary condition Eq.~(\ref{eq:mix_dyson_panel_bc}) can be written as
The boundary conditions in Eq.~(\ref{eq:mix_dyson_panel_bc}) can be reformulated using $P_n(\pm 1) = (\pm 1)^n$
\begin{align}
  \sum_n (-1)^n G^\ttau_{p,nm} & = i\xi (-1)^m G^M_m \, , & \text{for } p = 0 \, ,
  \nonumber \\
  \sum_n (-1)^n G^\ttau_{p,nm} & = \sum_n G^\ttau_{p-1,nm} \, , & \text{for } p > 0 \, .
  \label{eq:mix_dyson_panel_bc_legendre}
\end{align}
The action of the partial derivative $\partial_t$ on the real-time Legendre polynomial basis functions $P_n(\psi_p(t))$ of panel $p$ in Eq.~(\ref{eq:mix_dyson_panel}) is given by
\begin{equation}
  \partial_t P_n[\psi_p(t)]
  = \frac{2}{\Delta t} \partial_x P_n(x) \nonumber \\
  = \sum_{n'} D_{nn'} P_{n'}[\psi_p(t)] \, ,
\end{equation}
where $D_{nn'}$ is the upper triangular matrix \cite{Jie-Shen:2011uq}
\begin{align}
  \frac{\Delta t}{2} D_{nn'}
  \equiv
  \left\{
  \begin{array}{lr}
    2n' + 1, & 0 \le n' \le n, n'+n \textrm{ odd} \\
    0, & \textrm{elsewhere}
  \end{array}
  \right. \, .
  \label{eq:D_leg}
\end{align}

Using the expressions in Legendre coefficient space for the derivative [Eq.\ (\ref{eq:D_leg})], the history integral [Eq.\ (\ref{eq:I_hat_coeff})], and right-hand side [Eq.\ (\ref{eq:Qp_leg})], the panel Dyson equation~(\ref{eq:mix_dyson_panel}) can now be written entirely in Legendre coefficient space
\begin{multline}
  \sum_{n'} \left(
  iSD_{nn'} - F\delta_{nn'}
  - [\Sigma^R_0  \overset{<}{\ast}]_{nn'}
  \right) G^\ttau_{p,n'm}
  \\
  =
  Q^\ttau_{p,nm} + \mathcal{I}_{p, nm} + [\Sigma^R_p \overset{<}\ast G^\ttau_0]_{p,nm}
  \, .
  \label{eq:mix_dyson_panel_legendre}
\end{multline}
Equation (\ref{eq:mix_dyson_panel_legendre}) is a linear matrix equation of size $N_t$ for $G^\ttau_p$ on each panel $p$.

% --------------------------------------------------------------------
\begin{figure}
\includegraphics[scale=1]{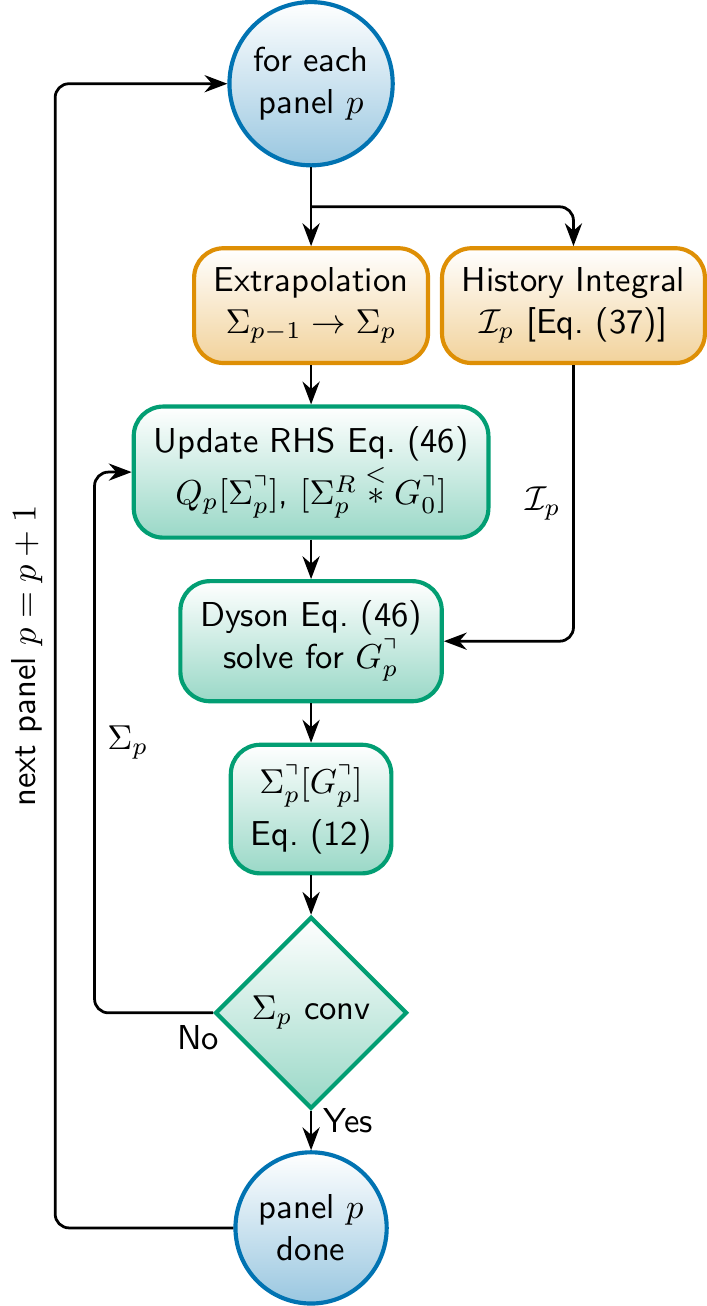} \\[-3mm]
\caption{\label{fig:sc} Schematic real-time panel propagation of $G^\ttau_p$ with self-energy self consistency imposed per panel. The history integral and extrapolation (orange boxes) is performed once per panel, while the Dyson equation and self-energy (green boxes) are iterated to self consistency in $\Sigma_p$.
} \end{figure}
% --------------------------------------------------------------------

% --------------------------------------------------------------------
\subsection{Time propagation of the real-time Dyson equation}
\label{sec:TimePropagationAlgo}
% --------------------------------------------------------------------

We summarize the algorithm for time propagation of the equilibrium real-time problem formulated in section \ref{sec:eqrtgf}.
The goal is to determine the mixed Green's function $G^\ttau(t, \tau)$ by self-consistently solving the real-time Dyson equation (\ref{eq:mix_dyson}) in combination with the self-energy relation $\Sigma^\ttau = \Sigma^\ttau[G\ttau]$ of Eq.\ (\ref{eq:sigma_func}).

The real-time panel subdivision of section \ref{sec:panel_exp} gives a real-time Dyson equation (\ref{eq:mix_dyson_panel}) that can be solved successively for each panel $p$, and its reformulation in Legendre coefficient space [Eq.\ (\ref{eq:mix_dyson_panel_legendre})] produces a linear system equation for $G^\ttau$.
The required calculational steps for the time propagation on panel $p$ are shown in Fig.\ \ref{fig:sc}.

For each panel $p$, the history integral $\mathcal{I}_p$ given by Eq.\ (\ref{eq:mix_dyson_panel_legendre_history}) is only computed once, since it depends on the Green's function $G^\ttau_{q}$ and self-energy $\Sigma^\ttau_{q}$ on earlier panels $q < p$.
For $p > 0$, an initial guess for the panel self energy $\Sigma_p$ is obtained by extrapolation of $\Sigma_{p-1}$ using linear prediction \cite{Barthel:2009kx}, in order to reduce the number of the self-energy self-consistent steps.
To emphasize that these two steps are only performed once per panel they are shown as orange boxes in Fig.\ \ref{fig:sc}.

The Dyson equation and self-energy self-consistency is performed by the steps represented as green boxes in Fig.\ \ref{fig:sc}. First, given the self-energy $\Sigma_p$ on the current panel $p$, the right-hand side terms in the Dyson equation (\ref{eq:mix_dyson_panel_legendre}), $Q_p[\Sigma^\ttau]$ and $[\Sigma^R_p \overset{<}\ast G^\ttau_0]$ are constructed. Then the panel Dyson equation (\ref{eq:mix_dyson_panel_legendre}) is solved for the panel Green's function $G^\ttau_p$, which in turn is used to compute the self-energy $\Sigma_p$ using Eq.\ (\ref{eq:sigma_func}). If the induced change in the self-energy $\Sigma_p$ is above a given threshold another self-energy self-consistent iteration is performed.
For the systems considered here, the relative change in the self-energy per iteration reaches machine precision in less than ten self-consistent iterations.
Once the self-energy is converged, the calculation for panel $p$ is complete and the time propagation proceeds to the next panel $p + 1$.

For long time simulations we observe spectral aliasing in the Legendre coefficients in imaginary time of $G^\ttau(t, \tau)$ (not shown). This phenomenon is well understood \cite{boyd::2000, PhysRevResearch.2.023068} and is resolved by using spectral-blocking in terms of Orzag's two-thirds rule \cite{boyd::2000}. In other words, the self-energy $\Sigma^\ttau(t, \tau)$ is evaluated on a denser collocation grid in imaginary time and only 2/3 of the resulting Legendre coefficients are used in the solution of the Dyson equation. This prevents the spectral aliasing and gives stable time-panel stepping.

% --------------------------------------------------------------------
\begin{figure}
\includegraphics[scale=1] {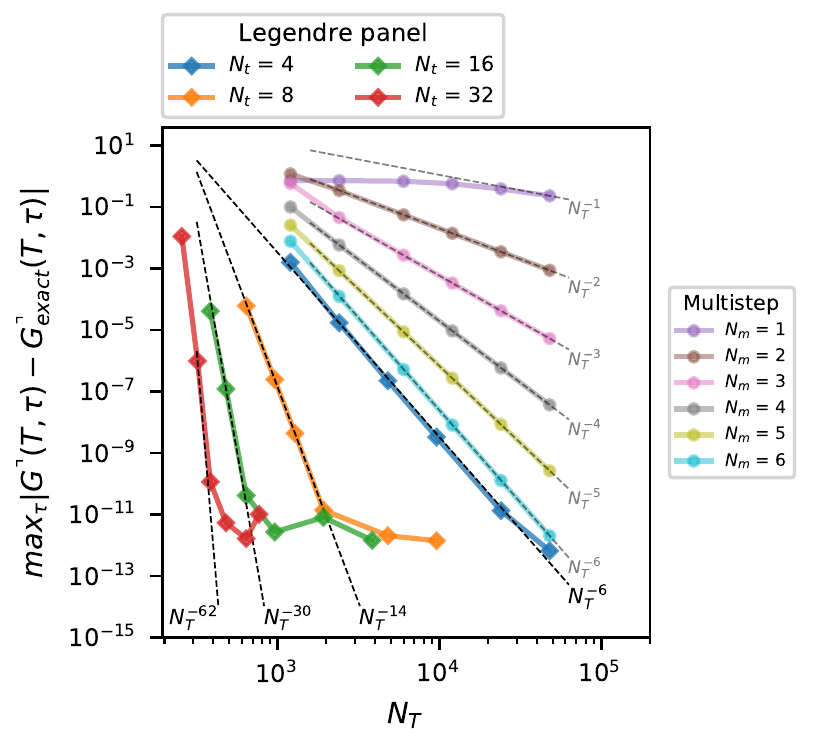} \\[-3mm]
\caption{
Error convergence of the equilibrium real-time Green's function $G_{11}^\ttau(T, \tau)$ at time $T = 48$ for a two level system, as a function of time discretization points $N_T$.
The real-time panel Dyson solver result for the panel expansion orders $N_t = 4$, $8$, $16$ and $32$ (diamonds), and the results of the equidistant multistep method of Ref.~\onlinecite{Schuler:2020aa} up to maximal order $N_m = 6$ (circles) are shown together with asymptotic convergence rates (dotted lines).
\label{fig:err_converge}
} \end{figure}
% --------------------------------------------------------------------
%
% --------------------------------------------------------------------
\section{Results -- Asymptotic convergence}
\label{sec:convergence}
% --------------------------------------------------------------------

To benchmark the convergence properties of the Legendre-panel based Dyson solver,
we use an analytically solvable two-level system
with energies $\epsilon_1$ and $\epsilon_2$ and hybridization $V$, giving
the matrix valued quadratic Hamiltonian
\begin{equation}
  h =
  \left[\begin{array}{cc}
      \epsilon_1 & V \\
      V & \epsilon_2
    \end{array}
    \right]
  \, .
\end{equation}
The matrix-valued contour Green's function $G_{ij}$ for this non-interacting system is given by
\begin{equation}
  (i \mathbf{1} \partial_z - h ) G = \delta_\mathcal{C}
  \label{eq:2lvl_matrix}
\end{equation}
and is solvable by explicit diagonalization.
The $G_{11}$ component of the Green's function also obeys the scalar Dyson equation of motion
\begin{equation}
  (i \partial_z - \epsilon_1 ) G_{11} - \int_\mathcal{C} d\bar{z} \Sigma(z, \bar{z}) G_{11}(\bar{z}, z) = \delta_\mathcal{C}
  \label{eq:2lvl_retarded}
\end{equation}
with the self-energy $\Sigma$ given by $\Sigma = V g_2 V$ with $g_2$ the solution of $(i\partial_z - \epsilon_2) g_2 = \delta_\mathcal{C}$. To derive Eq.\ (\ref{eq:2lvl_retarded}) from Eq.\ (\ref{eq:2lvl_matrix}) the inversion formulas for two-by-two block matrices can be used.

To benchmark our real-time panel Dyson equation solver we solve Eq.\ (\ref{eq:2lvl_retarded}) for $G^\ttau(t, \tau)$ in the equilibrium case $\epsilon_1=-1$, $\epsilon_2=5$, $V = 6$ at inverse temperature $\beta = 3$ and compare to the analytical solution obtained from Eq.\ (\ref{eq:2lvl_matrix}) at the final time $T=48$, see Fig.\ \ref{fig:err_converge}.
To be able to compare the results using different number of discretization points $N_t$ per panel, we study the error as a function of total number $N_T$ of time discretization points used, given by $N_T = N_p \cdot N_t$ where $N_p$ is the number of real-time panels.
For all $N_t$ we observe the asymptotic convergence rate
\begin{equation}
  \max_\tau | G^\ttau(T, \tau) - G^\ttau_{\text{exact}}(T, \tau) | \sim
  \mathcal{O}(N_T^{-2(N_t - 1)})
  \, .
  \label{eq:boundary_convergence}
\end{equation}
We note that there is no inherent limitation of the expansion order $N_t$, and a high order expansions like $N_t = 32$ gives an even higher order convergence rate $\sim \mathcal{O}(N_T^{-62})$.

We attribute the unexpected factor of two in the exponent of Eq.\ (\ref{eq:boundary_convergence}) to the superconvergence phenomenon \cite{Wahlbin:1995} present in the family of Galerkin methods of our Dyson solver [Eq.\ (\ref{eq:mix_dyson_panel_legendre})]. Numerical tests show that the convergence properties remain the same for $\Sigma = 0$ where Eq.\ (\ref{eq:mix_dyson_panel_legendre}) simplifies to a series of coupled first order initial value problems.
As shown in the literature \cite{doi:10.1137/0714015, douglas:1978, thomee:1980, Adjerid:2002wb}, and confirmed by our numerical tests, the high order superconvergence of Eq.\ (\ref{eq:boundary_convergence}) is only attained at the panel boundaries, while the remaining Legendre-Gauss-Lobatto collocation points (used for the self-energy evaluation)
converge as $\mathcal{O}(N_T^{-(N_t + 2)})$.
The observed superconvergence on the panel boundaries is beneficial since the initial value for each real-time panel [Eq.\ (\ref{eq:mix_dyson_panel_bc})] is known to high accuracy.

To put the convergence properties of our real-time panel Dyson solver in perspective,
we also solve Eq.\ (\ref{eq:2lvl_retarded}) using the state-of-the-art multistep method
for the real-time Dyson equation of Ref.\ \onlinecite{Schuler:2020aa}.
The multistep method uses an equidistant real-time discretization, the Gregory quadrature, and backward differentiation.
At order $N_m$ the asymptotic convergence of the multistep method is given by $\sim \mathcal{O}(N_T^{-N_m})$.
However, due to the inherent high-order instability of backward differentiation the order $N_m$ of the multistep method is limited to $N_m \le 6$ \cite{Schuler:2020aa}.
The convergence of the multistep method at all possible orders $N_m$ applied to the two level benchmark system is also shown in Fig.\ \ref{fig:err_converge}.

Comparing the performance of the two methods in Fig.~\ref{fig:err_converge}
explicitly shows the efficiency of high order polynomial panel expansions in real-time.
At equal orders $N_t$ and $N_m$ the asymptotic scaling of the multistep solver $N_T^{-N_m}$ is much slower than the $N_T^{-2(N_t-1)}$ rate of the real-time panel solver.
Hence, already at expansion order $N_t = 4$ the real-time panel solver (blue diamonds) has the same asymptotic error scaling $\mathcal{O}(N_T^{-6})$ as the maximum order $N_m = 6$ multistep method (cyan circles), see Fig.~\ref{fig:err_converge}.
In contrast to the multistep algorithm, the order $N_t$ of the Legendre-panel solver is not limited.
Going to high polynomial order gives a dramatic reduction in the total number of time discretization points $N_T$ required to reach high accuracy.
For example, reaching an accuracy of $10^{-11}$ using expansion order $N_t = 4$ requires $N_T \sim 10^4$ points, while using expansion order $N_t = 32$ reduces the required number of required real-time points to $N_T \sim 10^2$, i.e.\ by almost two orders of magnitude.

Thus, for a fixed final time $T$ and accuracy, the number of real-time discretization points $N_T$ required to store the equilibrium real-time Green's function can be drastically reduced when using the high order real-time panel Dyson solver.
This is an important advance since calculations in general are memory limited, in particular when using the multistep method.
Using the high-order real-time panel expansion will therefore enable the study of both larger systems and longer simulation times.

% --------------------------------------------------------------------
\subsection{Computational complexity}
\label{sec:complexity}
% --------------------------------------------------------------------
%
% --------------------------------------------------------------------
%
\begin{table}
\begin{tabularx}{1.0\columnwidth}
  {>{\raggedleft\arraybackslash}X||c|>{\raggedright\arraybackslash}X r}
  &  Multistep & Legendre-Panel & Eq. \\
  \hline
  && \\[-0.3cm]
  Linear system & $N_T$ & $N_T \cdot N_t^2$ \footnote{$N_T \cdot N_t$ with iterative linear solver.} & (\ref{eq:mix_dyson_panel_legendre}) \\
  History integral & $N_T^2$ &  $N_T^2$ & (\ref{eq:mix_dyson_panel_legendre_history})
\end{tabularx}
\caption{
  Asymptotic computational complexity
  as a function of time discretization points $N_T$ for the multistep method of Ref.\ \onlinecite{Schuler:2020aa} and the real-time panel method at panel expansion order $N_t$
  \label{tab:scaling} }
\end{table}
%
% --------------------------------------------------------------------
%
In the previous section, it was shown that high-order real-time panel expansions reduce the required number of time discretization points by orders of magnitude for a fixed level of accuracy, as compared to the state-of-the-art multistep method of Ref.\ \onlinecite{Schuler:2020aa}.
This enhanced performance comes at the price of a moderate increase of computational complexity in the linear system solver step, see Tab.\ \ref{tab:scaling}.

The main difference between the multistep solver and the real-time panel solver is that the panel based approach requires solving Eq.~(\ref{eq:mix_dyson_panel_legendre}) for all time points within a panel at once.
This amounts to solving a per-panel linear system with a naive cubic scaling $\mathcal{O}(N_t^3)$, producing the extra prefactor $N_t^2$ in the computational complexity of the linear system in Tab.\ \ref{tab:scaling}.
Using a preconditioned iterative linear solver may reduce this by one factor of $N_t$ and is an interesting venue for further research.
Even though this step of the Dyson equation has a higher computational complexity, this is not an issue when taking into account the reduction of $N_T$ enabled by the high-order expansion. Furthermore, the solution of the linear system is in fact not the computational complexity bottleneck of the Dyson solver.

The main computational bottleneck of the Dyson equation is the calculation of the history integral [Eq.\ (\ref{eq:mix_dyson_panel_legendre_history})].
In the direct multistep method the history integral evaluation scales quadratically as $\mathcal{O}(N_T^2)$, and the real-time panel history integral in Eq.\ (\ref{eq:mix_dyson_panel_legendre_history}) retains the same scaling $\mathcal{O}(N_T^2) = \mathcal{O}(N_p^2 \cdot N_t^2)$ by using the recursive Legendre convolution algorithm \cite{Hale:2014ab}.
However, in the special case of equilibrium real-time it was recently shown that the scaling of the history integral can be reduced to quasi-linear scaling \cite{Kaye:2021aa}. The generalization of this approach to the real-time panel expansion is another promising direction for further research.

Potential computational complexity gains from the linear system and the history integral aside, the real-time panel Dyson solver algorithm presented here is already competitive for memory-limited problems.
By extending the range of applicability of real-time propagation via the drastically lower number of discretization points $N_T$ needed for a given accuracy, see Sec.\ \ref{sec:convergence}.
The same compactness property also makes the generalization of the real-time panel discretization from equilibrium real-time to non-equilibrium real-time propagation an interesting direction of further research.

% --------------------------------------------------------------------

% --------------------------------------------------------------------
\section{Results -- Application to molecules}\label{sect: application}
\label{sec:GF2}
% --------------------------------------------------------------------

As a proof-of-concept application of the equilibrium real-time Dyson equation solver, we solve the real-time propagation of the mixed Green's function $G^\ttau$ for several molecules, using dressed second order perturbation theory (GF2) \cite{PhysRevB.63.075112, Dahlen:2005aa, Phillips:2014aa, Phillips:2015ab,
  Kananenka:2016ut, Kananenka:2016ab, :/content/aip/journal/jcp/144/5/10.1063/1.4940900, :/content/aip/journal/jcp/145/20/10.1063/1.4967449, PhysRevB.100.085112}.
%
%The starting point of the real-time propagation is chosen to be the self-consistent imaginary-time solution to preserve the time-translation invariant.
%
The calculations are using standard Gaussian basis functions and matrix elements from the quantum chemistry code pySCF \cite{Sun:2018aa, Sun:2020ui}, and the initial condition $G^M$ in Eq.\ \ref{eq:mix_initial_cond} is obtained using our in-house GF2 code implementing the Legendre spectral algorithm detailed in Ref.\ \onlinecite{doi:10.1063/5.0003145}.

Performing explicit time-propagation enables us to avoid the ill posed analytical continuation problem \cite{Jarrell:1996fj, PhysRevLett.126.056402, PhysRevB.104.165111} by computing the real-frequency spectral function $A(\omega)$ as a direct Fourier transform of $G^R$, see Eqs.\ (\ref{eq:A_omega_from_GR}-\ref{eq:GR_from_Gtv}).
The spectral function is in turn used to determine the electron affinity (EA) and the ionization potential (IP) given by the first excitation peaks in $A(\omega)$ above and below zero frequency.

For the small molecules H$_2$ and LiH we compute the total energy, the spectral function, IP and EA, as a function of the inter-atomic separation $r$ and compare with standard quantum chemistry methods like Hartree-Fock (HF), M{\o}ller-Pleset perturbation theory (MP2), Coupled Cluster Singles Doubles (CCSD), and Full Configuration Interaction (FCI). In particular, for HF the spectra is computed using Koopmans Theorem (HF-KT) \cite{Koopmans:1934wi}, and for CCSD we use the Equation Of Motion (CCSD-EOM) technique \cite{Stanton:1996vw, Saeh:1999tb}. We also investigate the approximated GF2 spectral function obtained from the Extended Koopmans Theorem (GF2-EKT) \cite{Smith:1975ww, Day:1975wm, Morrison:1975ux, Ellenbogen:1977uh, Chipman:1977va, Vanfleteren:2009ul} and compare to the exact GF2 spectra obtained from the time evolution (GF2-RT).

For the larger molecule Benzoquinone (C$_6$H$_4$O$_2$), out of reach for the methods CCSD and FCI, we compare with HF and the AGF2 method \cite{Backhouse:2020aa, Backhouse:2020ab} which is an approximation to self-consistent GF2.

The reference HF, CCSD, CCSD-EOM, and AGF2 calculations are performed using pySCF \cite{Sun:2018aa, Sun:2020ui} while the FCI spectral function is computed using EDLib \cite{Iskakov:2018aa}.

% --------------------------------------------------------------------
\subsection{Real-time second order self-energy}
\label{sec:GF2_Sigma}
% --------------------------------------------------------------------

Within the dressed second order self-energy approximation (GF2) \cite{PhysRevB.63.075112, Dahlen:2005aa, Phillips:2014aa} the mixed self-energy $\Sigma^\ttau$ is given by the direct product of three Green's functions
\begin{multline}
    \Sigma^\ttau_{ij}(t, \tau) = v_{ilnp}(2v_{jkqm} - v_{jqkm}) \\
    \times G^\ttau_{lk}(t, \tau) G^\ttau_{pq}(t, \tau) G^{\ttau *}_{nm}(t, \beta - \tau) \, ,
    \label{eq:mix_sigma_gf2}
\end{multline}
where $v_{ijkl}$ is the electron-electron Coulomb repulsion integral \cite{Stefanucci:2013oq}. The analogous expression for $\Sigma^M$ in Ref.\ \onlinecite{doi:10.1063/5.0003145} is obtained using the initial condition in Eq.\ (\ref{eq:mix_initial_cond}), and the retarded-self energy $\Sigma^R$ is directly given by $\Sigma^\ttau$ using Eq.\ (\ref{eq:sigma_r}).

In our GF2 calculations the self-energy calculation step using Eq.\ (\ref{eq:sigma_func}) in the real-time panel time propagation algorithm of Fig.\ \ref{fig:sc} is replaced by the GF2 self-energy expression [Eq.\ (\ref{eq:mix_sigma_gf2})].

% --------------------------------------------------------------------
\begin{figure}
\includegraphics[scale=1]{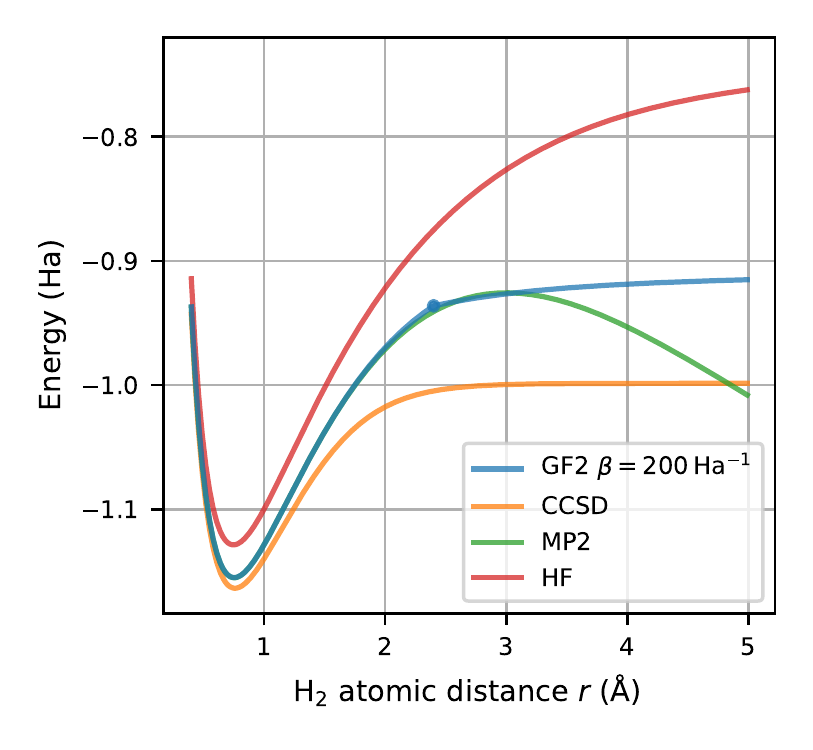}
\caption{\label{fig:H2_stretch}Total energy $E$ of H$_2$ in the cc-pVDZ basis as a function of inter-atomic distance $r$ using GF2, CCSD, MP2, and HF. Note that for H$_2$ with two electrons CCSD is exact.
} \end{figure}
% --------------------------------------------------------------------

% --------------------------------------------------------------------
\begin{figure}
\includegraphics[scale=1]{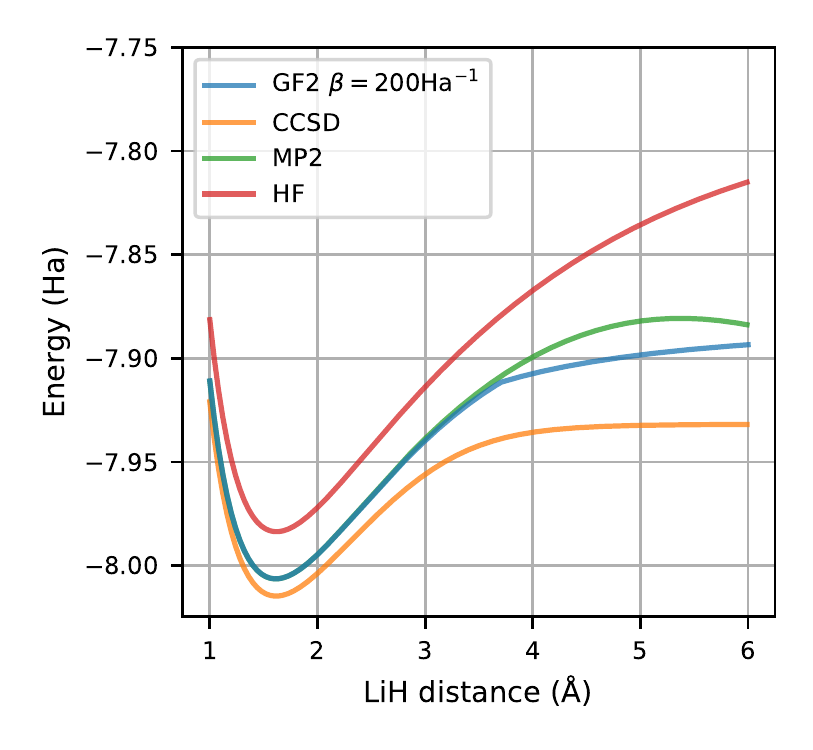}
\caption{\label{fig:LiH_stretch}Total energy $E$ of LiH in the cc-pVDZ basis as a function of inter-atomic distance $r$ using GF2, CCSD, MP2, and HF.
} \end{figure}
% --------------------------------------------------------------------

% --------------------------------------------------------------------
\subsection{Small molecules: H$_2$ and LiH}
% --------------------------------------------------------------------
%
For the small molecules H$_2$ and LiH we first compute total energy as a function of inter-atomic distance $r$ in the cc-pVDZ basis and compare GF2 with HF, MP2, and CCSD, in Figs.\ \ref{fig:H2_stretch} and \ref{fig:LiH_stretch}.
The GF2 total energy $E$ is given by
\begin{align}
E = \frac{1}{2} \text{Tr}[(h + F)P] + \text{Tr}[\Sigma * G] + E^{(nn)}
    \, ,
    \label{eq:E_GF2}
\end{align}
where $E^{(nn)}$ is the nuclei-nuclei Coulomb energy and $P$ is the density matrix given by $P = -2 G^M(\beta)$.
The difference between MP2 and GF2 is the self-consistency, and our results reproduce the well-known observation \cite{Phillips:2014aa, Backhouse:2020aa} that the divergence of the total energy of MP2 at large $r$ is not present in GF2, where the total energy instead levels out for large $r$, see Figs.\ \ref{fig:H2_stretch} and \ref{fig:LiH_stretch}.

In the intermediate range of inter-atomic separation $r$ there are two self-consistent GF2 solutions, which are adiabatically connected to the low and high $r$ regimes.
The total energy of the two solutions cross at intermediate values of $r$ \cite{Phillips:2014aa} and the curves in
Figs.\ \ref{fig:H2_stretch} and \ref{fig:LiH_stretch} show the lowest energy solution.
The coexistence of multiple solutions in dressed perturbation theory is an active field of research \cite{1367-2630-17-9-093045, PhysRevB.94.235108, PhysRevB.98.235107, Kozik:2015aa, Rossi:2015vy, PhysRevLett.119.056402, PhysRevLett.125.196403, Iskakov:2022}.

At the equilibrium distance $r_0$ the total energy $E(r)$ is minimized, and for H$_2$ and LiH we observe that the GF2 total equilibrium energy does not improve on the MP2 result relative to the CCSD result, which is exact for H$_2$ with only two electrons.
%
%This is not generic, since GF2 performs significantly better than MP2 for $E(r_0)$ (relative to CCSD) in other systems like He$_2$ \cite{doi:10.1063/5.0003145}.
%
This is not generic, since GF2 performs significantly better than MP2 (relative to CCSD) in other cases. One example is the dissociation energy of He$_2$ \cite{doi:10.1063/5.0003145}, see Appendix \ref{app:He2} for a comparison of the spectral functions.

% --------------------------------------------------------------------
\begin{figure*}[tbh]
\includegraphics[scale=1]{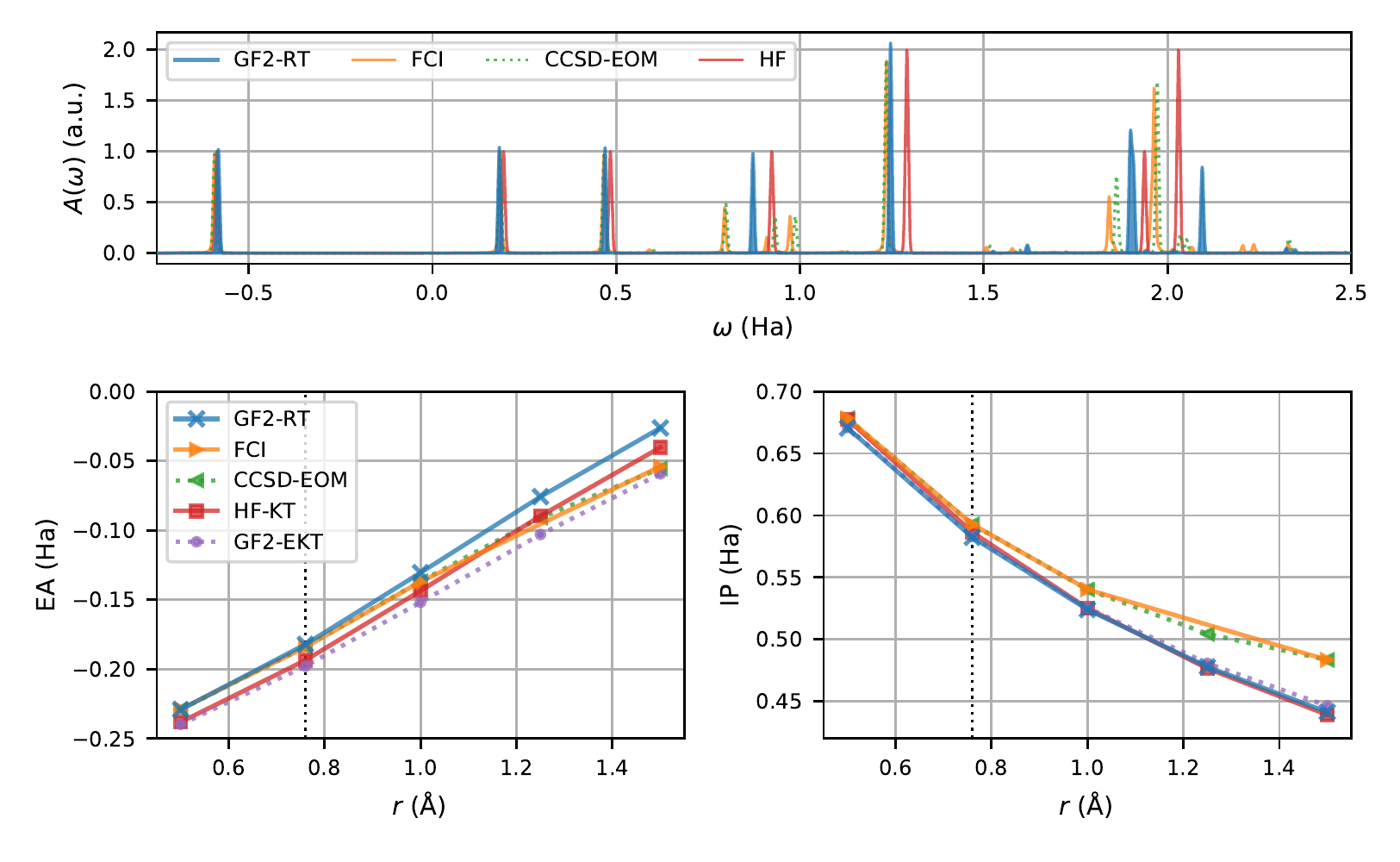} \\[-8mm]
\caption{\label{fig:H2_spectra}Upper panel: Spectral function of H$_2$ at equilibrium H-H distance $r_0 = 0.76\,$Å using the cc-pVDZ basis. The GF2 result is compared to Hartree-Fock (HF) and the Coupled Cluster Singles-Doubles Equation of motion (CCSD-EOM). Lower left panel: Electron affinity (EA) as a function of $r$. Lower right panel: Ionization potential (IP) as a function of $r$.
} \end{figure*}
% --------------------------------------------------------------------

% --------------------------------------------------------------------
\begin{figure*}[tbh]
\includegraphics[scale=1]{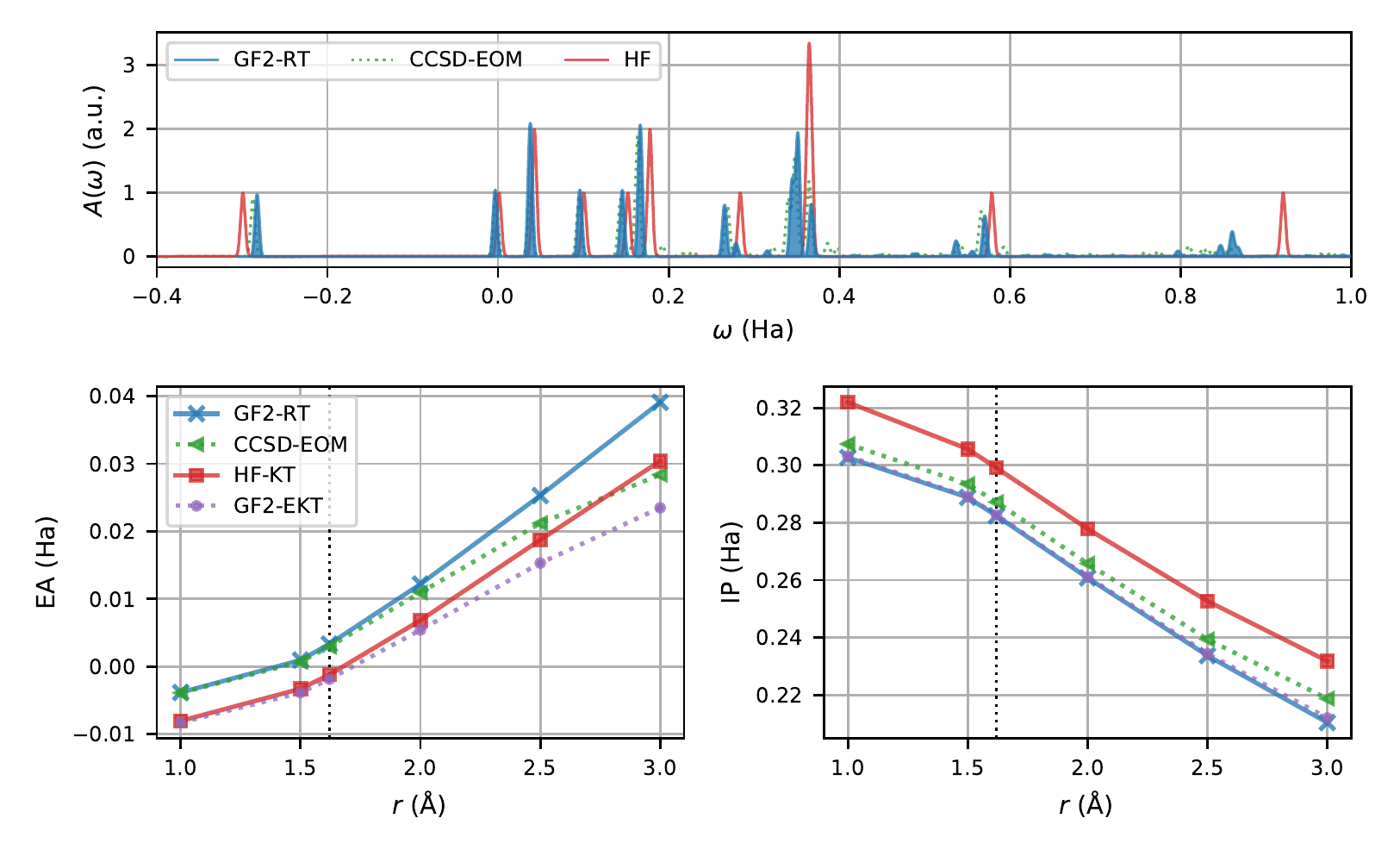} \\[-8mm]
\caption{\label{fig:LiH_spectra}Upper panel: Spectral function of LiH at equilibrium Li-H distance $r_0 = 1.62\,$Å, $\beta = 200$ using the cc-pVDZ basis. The GF2 result is compared to Hartree-Fock (HF) and the Coupled Cluster Singles-Doubles Equation of motion (CCSD-EOM). Lower left panel: Electron affinity (EA) as a function of $r$. Lower right panel: Ionization potential (IP) as a function of $r$.
} \end{figure*}
% --------------------------------------------------------------------

% --------------------------------------------------------------------
\subsubsection{Equilibrium spectral function}
% --------------------------------------------------------------------

To determine the equilibrium spectral function $A(\omega)$ we perform GF2 equilibrium time-propagation of $G^\ttau$ using the real-time panel algorithm of section \ref{sec:TimePropagationAlgo} and the GF2 self-energy in Eq.\ (\ref{eq:mix_sigma_gf2}).
From $G^\ttau$ the retarded Green's function $G^R$ is obtained using Eq.\ (\ref{eq:GR_from_Gtv}) that in turn gives the spectral function as
\begin{equation}
  A(\omega) = - \frac{1}{\pi} \text{Im}( \text{Tr}[ S G^R(\omega) ] )
  \label{eq:SGR_A}
\end{equation}
where $S$ is the overlap matrix.

For H$_2$ and LiH the time propagation is performed using real-time panels with $16^\text{th}$ order Legendre expansions ($N_t = 16$) yielding floating point accuracy for the panel time step sizes $\Delta t \approx 19 \, \text{as}$ ($0.8\,\text{Ha}^{-1}$) and $15 \, \text{as}$ ($0.6\,\text{Ha}^{-1}$), respectively.
The propagation times are $t_{max} \approx 19 \, \text{fs}$ ($800\,\text{Ha}^{-1}$) and $29 \, \text{fs}$ ($1200\,\text{Ha}^{-1}$), giving the frequency resolutions $\Delta \omega = \pi / t_{max} \approx 0.004 \, \text{Ha}$ and $0.003\,\text{Ha}$, for H$_2$ and LiH respectively. The resulting spectral functions for H$_2$ and LiH at the equilibrium atomic distance are shown in the upper panels of Fig.~\ref{fig:H2_spectra} and Fig.~\ref{fig:LiH_spectra}, together with the HF-KT, CCSD-EOM, and FCI spectra at the same energy resolution.

To better reveal many-body effects the spectral function $A(\omega)$ is scaled with $\sqrt{2\pi}/(\Delta \omega)$ causing a single-particle-state peak with a Gaussian broadening of $\sigma = \Delta \omega$ to have unit height. With this scaling the individual peaks in the HF-KT spectra all have integer height, while many-body correlations drive peak height renormalization (away from integer values) for the methods GF2, CCSD-EOM and FCI.

Comparing the GF2 spectral function for H$_2$ in Fig.\ \ref{fig:H2_spectra} with HF, CCSD-EOM, and the exact FCI results, we see that GF2 is an overall improvement comparing to HF.
The position of the occupied state at $\omega \approx -0.58\,$Ha is roughly the same for all methods, however, GF2 is actually slightly worse than HF when compared to the exact FCI result.
For all other spectral features, GF2 is an improvement compared to HF. In GF2 the two first peaks at positive frequencies are shifted down relative to HF, in agreement with FCI. For the higher spectral features the frequency moments of GF2 are improved over HF, while the peak structure differs from FCI.
We also note that CCSD-EOM agrees remarkably well with the exact FCI spectra. Thus, for LiH where FCI is out of reach we will use CCSD-EOM as the base line comparison for GF2.

For LiH the GF2 spectra agree even better with the CCSD-EOM spectra as compared to HF. Relative to the HF spectra, the first peak at negative frequencies is shifted up in frequency, while the peaks at positive frequencies are shifted down, all in agreement with CCSD-EOM. While the low frequency peak heights are only weakly renormalized, we also note that GF2 correctly captures the strong renormalization of the spectral feature at $\omega \approx 0.275\,$Ha.

The good agreement in equilibrium spectra between GF2 and CCSD-EOM (and FCI) is promising, in particular for the application of GF2 to investigate non-linear processes in molecular systems out of equilibrium \cite{Leeuwen_2006, Stan_2006, PhysRevA.91.033416, PhysRevA.92.033419}.
However, for finite systems and small basis sets, care must be taken with regard to damping effects from infinite diagram resummation as seen in simple model systems like Hubbard clusters \cite{PhysRevLett.103.176404, PhysRevB.82.155108}.

% --------------------------------------------------------------------
\begin{figure*}[t]
\includegraphics[scale=1]{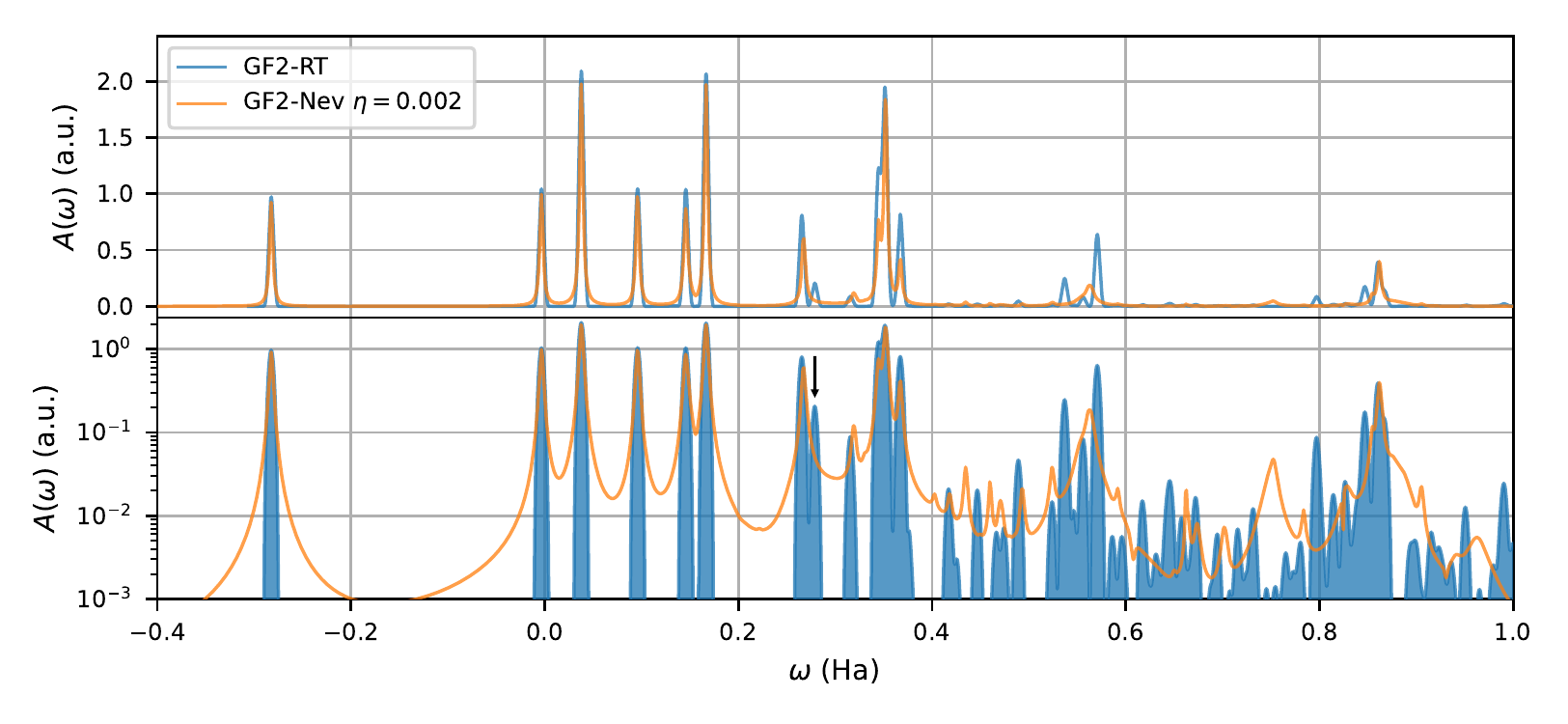}
\caption{\label{fig:LiH_Nevspectra}
  GF2 spectral function (blue) from real-time propagation compared to the spectral function obtained by analytical continuation of the GF2 solution in imaginary time using the Nevanlinna method (orange) \cite{PhysRevLett.126.056402}, for LiH in the cc-pVDZ basis at $\beta = 200$ and equilibrium inter-atomic distance $r_0 = 1.62\,$Å, on a linear scale (upper panel) and logarithmic y-axis (lower panel).
  The spectral functions are scaled so that a non-degenerate single-particle state has a peak height of unity.
} \end{figure*}
% --------------------------------------------------------------------

% --------------------------------------------------------------------
\subsubsection{Comparison with analytical continuation}
% --------------------------------------------------------------------

Within the GF2 self-energy approximation the spectral function $A(\omega)$ is obtained from real-time propagation at energy resolution $\Delta \omega = \pi / t_{max}$.
Having the spectral function enables us to benchmark the Nevanlinna analytical continuation method \cite{PhysRevLett.126.056402}.
Analytical continuation solves the ill-posed inverse problem of determining an approximate spectral function using only the imaginary time Green's function $G^M(\tau)$ \cite{Jarrell:1996fj}.

In Fig.~\ref{fig:LiH_Nevspectra} the GF2 spectral function for LiH (at energy resolution $ \approx 0.003\,\text{Ha}$) is compared with the Nevanlinna spectral function.
The Nevanlinna calculation was performed for each diagonal component of the $S G^M(\tau)$ product, c.f.\ Eq.\ (\ref{eq:SGR_A}), using 225 positive Legendre sparse-sampling Matsubara frequencies (App.~\ref{app:SparseSampling}) and 25 Hardy basis functions (see Ref.~\onlinecite{PhysRevLett.126.056402}), evaluated $0.002\,$Ha above the real-frequency axis.
As seen in in Fig.~\ref{fig:LiH_Nevspectra}, peaks up to $\approx 0.20\,$Ha are well captured by the Nevanlinna method.
However, some of the higher energy correlated resonances are missed or smeared out, such as the one at $\approx 0.275\,$Ha (black arrow).

We stress that the equilibrium real-time propagation method proposed in this manuscript eliminates the need for analytical continuation.

% --------------------------------------------------------------------
\subsubsection{Ionization potential and electron affinity}
% --------------------------------------------------------------------

At positive frequencies $\omega > 0$ the spectral function $A(\omega)$ describes electron addition excitations, while negative frequencies $\omega < 0$ corresponds to electron removal excitations.
Hence, the minimal energy for electron removal, the ionization potential (IP) and the minimal energy for electron addition, the electron affinity (EA), are given by the first peak in $A(\omega)$ below and above $\omega=0$, respectively.
To investigate how GF2 performs both in the weakly and strongly correlated regimes, we study the IP and EA as a function of inter-atomic distance $r$ for H$_2$ and LiH.
For large inter-atomic separations $r \gg r_0$ the kinetic overlaps become exponentially small while the long range Coulomb interaction varies weakly.
The GF2 result is compared to the HF-KT and the CCSD-EOM results, as well as the exact FCI result in the case of H$_2$.

For H$_2$ the IP and EA are shown in Fig.~\ref{fig:H2_spectra} as a function of $r$.
The overall performance of GF2 relative to the exact FCI result is better in the weakly correlated regime $r \lesssim r_0$, compared to the strongly correlated regime $r \gg r_0$.
The GF2 behavior relative to HF, however, is different for the IP and EA even in the weakly correlated regime. For the EA, GF2 constitutes a drastic improvement over HF, while for the IP, GF2 largely follows the HF result.
We note that the exact FCI result is closely followed by CCSD-EOM, which is used as baseline comparison for LiH.
The IP and EA for LiH are shown in Fig.~\ref{fig:LiH_spectra}.
For both IP and EA we find that GF2 performs significantly better than HF relative to the CCSD-EOM result.
However, the GF2 behavior as a function of $r$ differs between IP and EA when entering the strongly correlated regime. The EA deviates from CCSD-EOM while the IP follows the $r$ dependence of CCSD-EOM with small offset.

In the light of the perturbation expansion order, the observed progression from HF to GF2 shows that, going from the first order dressed perturbation expansion of HF to the second order dressed perturbation expansion GF2, improves the excitation spectra in the weakly correlated regime. However, in the strongly correlated regime, with larger interaction to kinetic overlap ratios, also the GF2 second order perturbation expansion does not suffice.
Hence, GF2 is probably not well suited for studying phenomena in the $r \gg r_0$ regime like dynamical atomic dissociation. However, it is a promising level of approximation to study phenomena at $r \sim r_0$, like non-linear optical-vibronic dynamics, terahertz response, and high harmonic generation \cite{RevModPhys.81.163}.

Finally we connect to previous diagrammatic perturbation theory works computing IP and EA from the imaginary time Green's function $G^M(\tau)$ using the extended Koopmans theorem (EKT) \cite{Smith:1975ww, Day:1975wm, Morrison:1975ux, Ellenbogen:1977uh, Chipman:1977va, Vanfleteren:2009ul}.
Within EKT, electron addition and removal energies are computed from a generalized eigenvalue problem constructed from $G^M(\tau)$ and $\partial_\tau G^M(\tau)$ at $\tau = 0^{\pm}$, see Appendix \ref{app:EKT} for details. It has been used to compute IP and EA both from GW \cite{Stan:2006vc} and GF2 \cite{Dahlen:2005aa, Welden:2015aa, PhysRevB.97.115164} imaginary time calculations.
However, how accurate the EKT approach is relative to the actual IP and EA of the spectral function $A(\omega)$ has not been investigated.

The real-time propagation approach presented here directly gives the spectral function $A(\omega)$ and alleviates the need for using EKT to compute the IP and EA. However, it also makes it possible to investigate the accuracy of EKT by direct comparison to the exact spectral-function derived IP and EA. The real-time GF2-RT and the GF2-EKT results for the IP and EA are shown for H$_2$ and LiH in Fig.~\ref{fig:H2_spectra} and Fig.~\ref{fig:LiH_spectra}, respectively. In both cases the EA from GF2-EKT fails to reproduce the GF2-RT result, instead the EKT calculations give EAs that match the HF results for $r \lesssim r_0$. These results raise serious concerns regarding the use of EKT for computing EAs in GF2.

% --------------------------------------------------------------------

% --------------------------------------------------------------------
\subsection{Intermediate size molecule: Benzoquinone C$_6$H$_4$O$_2$}
% --------------------------------------------------------------------

% --------------------------------------------------------------------
\begin{figure*}
\includegraphics[scale=1]{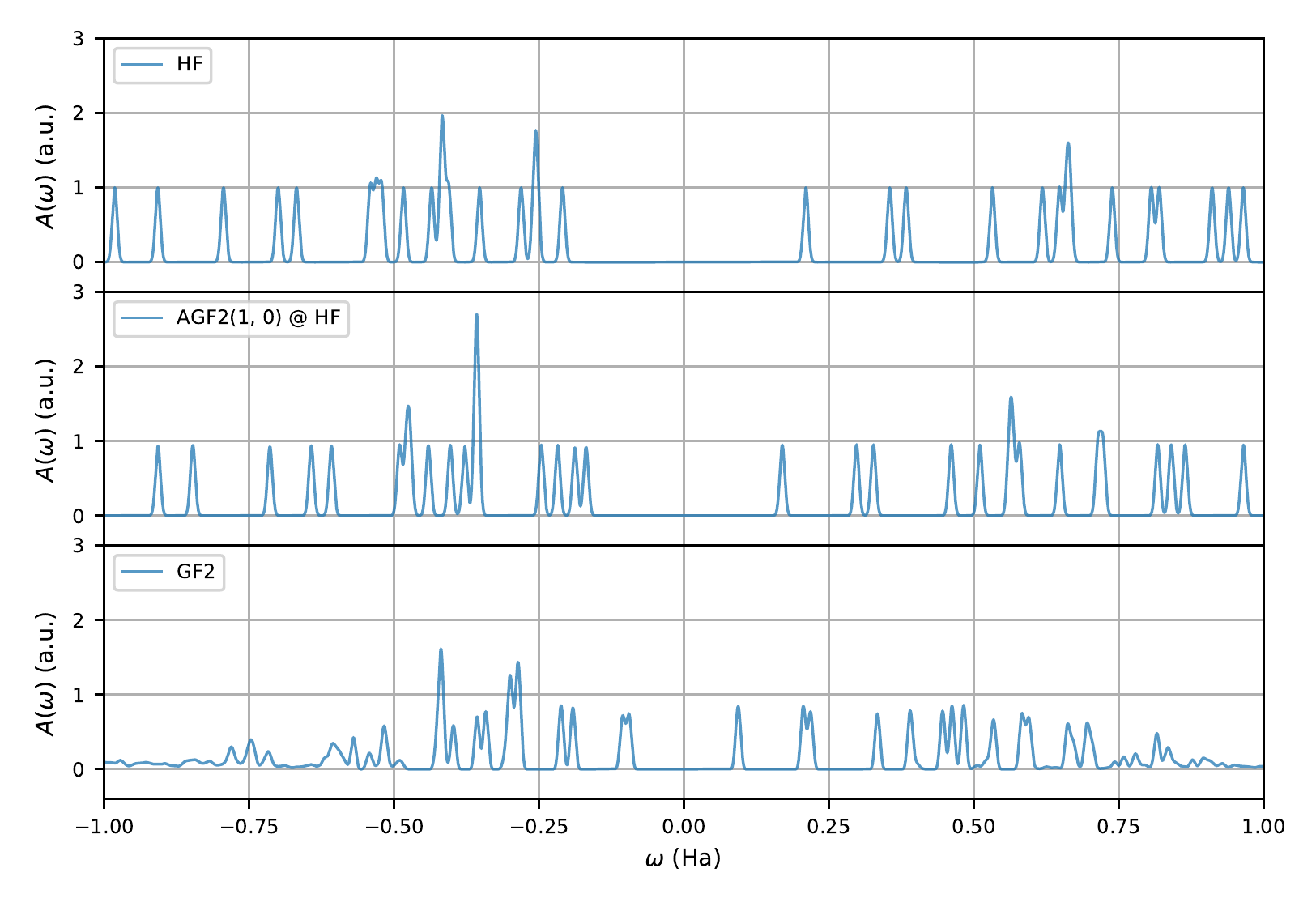}
\caption{\label{fig:C6H4O2_spectra} Spectral function of Benzoquinone (C$_6$H$_4$O$_2$) in the STO-3g basis from HF (upper panel), AGF2(1, 0)@HF (middle panel), and GF2 (lower panel).
} \end{figure*}
% --------------------------------------------------------------------

To explore the solver in a regime that is not otherwise accessible, we compute the spectral function of the Benzoquinone molecule (C$_6$H$_4$O$_2$) in a minimalistic STO-3g basis (44 basis functions), with optimized MP2 geometry \cite{Johnson:2020aa}.
A previous density functional study has shown that the HOMO-LUMO gap of Benzoquinone can not be described by ab initio density functionals like PBE \cite{Gallandi:2016vn}, while HF overestimates the gap.
However, a recent study \cite{Backhouse:2020ab} have shown that a self consistent approximate formulation of GF2, called the auxiliary second-order Green's function perturbation theory (AGF2), is able to describe the experimental gap.

For the real-time propagation a $16^\text{th}$ order real-time panel expansion was used with panel time step size $\Delta t \approx 7.3 \, \text{as}$ ($0.3\,\text{Ha}^{-1}$) and a total propagation time of $t_{max} \approx 18 \, \text{fs}$ ($750\,\text{Ha}^{-1}$).
The minimal STO-3g basis prevents direct comparison with experiments, and we compare to AGF2 and HF in this basis.
The total memory foot-print of the calculation is on the order of 500 GB.
The molecular point group symmetry is also used to speed up the GF2 self-energy evaluation.

Figure~\ref{fig:C6H4O2_spectra} shows the GF2 spectral function of Benzoquinone together with the corresponding results from HF and AGF2(1,0)@HF \footnote{See Refs.\ \onlinecite{Backhouse:2020aa, Backhouse:2020ab} for details on the partial selfconsistency notation: AGF2(X,Y).}.
The corresponding HOMO-LUMO gaps listed in Tab.~\ref{tab:HOMOLUMO}, shows that, going from first order HF, through the approximate second order AGF2(1,0)@HF result, to the full second order self-consistent GF2 result, yields a decreasing HOMO-LUMO gap.
Accounting for the aug-cc-pVDZ results for HF and AGF2(1,0)@HF from Ref.\ \onlinecite{Backhouse:2020ab}, see Tab.\ \ref{tab:HOMOLUMO}, the experimental HOMO-LUMO gap of $0.299\,$Ha \cite{Dougherty:1977va, Fu:2011ws} is likely to be underestimated by GF2 also in the larger aug-cc-pVDZ basis.

Another distinct feature of the full GF2 spectral function is the large degree of quasi-particle renormalization, as measured in terms of deviation from unit height in the spectral function, see Fig.\ \ref{fig:C6H4O2_spectra}.
This is to be compared with HF where all individual excitations come with unit height and the partial self-consistent AGF2 that only yields a small frequency-independent renormalization. The GF2 spectral function, on the other hand, displays peak-height renormalizations of the order 10-20\% for the HOMO and LUMO peaks and even a loss of coherence for the spectra at larger frequencies.

% --------------------------------------------------------------------
%
\begin{table}
\begin{tabularx}{1.0\columnwidth}
  {c|YY}
  \backslashbox{Theory}{Basis} & STO-3g & aug-cc-pVDZ \\
  \hline
  && \\[-0.3cm]
  HF & 0.420 Ha & 0.410 Ha\\
  AGF2(1,0)@HF & 0.338 Ha & 0.372 Ha \\
  GF2 & 0.189 Ha & - \\
  \hline \\[-2mm]
  Exp \cite{Dougherty:1977va, Fu:2011ws} & 0.299 Ha
\end{tabularx}
\caption{HOMO-LUMO gap of Benzoquinone (C$_6$H$_4$O$_2$) at different levels of theory in the STO-3g basis (this work) and the aug-cc-pVDZ basis from Ref.~\onlinecite{Backhouse:2020ab}.\label{tab:HOMOLUMO}}
\end{table}
%
% --------------------------------------------------------------------

% --------------------------------------------------------------------
\section{Conclusion and outlook}\label{sec:conclusion}
% -------------------------------------------------------------------

We present a panel discretization of the real-time axis for contour Green's functions using a piece-wise high-order orthogonal Legendre polynomial expansion.
Using this expansion to represent the mixed Green's function $G^\ttau(t, \tau)$ [Eq.\ (\ref{eq:gttau_product_basis})], we show a drastic reduction of the required number of discretization points needed to reach fixed accuracy, as compared to state-of-the-art multistep methods \cite{Schuler:2020aa}.

This result is achieved using a superconvergent \cite{Wahlbin:1995} algorithm for solving the equilibrium real-time Dyson equation of motion which we describe in detail.
The algorithm uses the Legendre spectral method \cite{Jie-Shen:2011uq} in combination with a recursive algorithm for Legendre convolution \cite{Hale:2014ab}.
The superconvergence \cite{Wahlbin:1995, doi:10.1137/0714015, douglas:1978, thomee:1980, Adjerid:2002wb} gives a panel-boundary error scaling $\mathcal{O}(N_T^{-2(N_t - 1)})$ for the total number of real-time discretization points $N_T$ and $N_t$ points per panel.
When combined with analytical self-energy approximations like GF2 \cite{PhysRevB.63.075112, Dahlen:2005aa, Phillips:2014aa, Phillips:2015ab, Kananenka:2016ut, Kananenka:2016ab, :/content/aip/journal/jcp/144/5/10.1063/1.4940900, :/content/aip/journal/jcp/145/20/10.1063/1.4967449, PhysRevB.100.085112}, the equilibrium real-time propagation of $G^\ttau(t, \tau)$ can be used to determine the real-frequency spectral function to an accuracy $\Delta \omega$ only limited by the total simulation time $t_{max}$, $\Delta \omega \approx \pi / t_{max}$.

As proof-of-concept, we compute the molecular spectral function of H$_2$, LiH, and C$_6$H$_4$O$_2$ by equilibrium real-time evolution of $G^\ttau(t,\tau)$ on the level of dressed second-order Green's function perturbation theory (GF2) \cite{PhysRevB.63.075112, Dahlen:2005aa, Phillips:2014aa, Phillips:2015ab}, and compare to standard quantum chemistry methods and the approximated auxiliary GF2 method \cite{Backhouse:2020aa, Backhouse:2020ab}.
Having the GF2 spectral function (up to resolution $\Delta \omega$) also enables stringent benchmarking of analytical continuation \cite{Jarrell:1996fj}, and we present a comparison of the Nevanlinna method \cite{PhysRevLett.126.056402} on LiH.

Our molecular GF2 calculations establish the applicability of the high-order expansion methods for equilibrium real-time evolution of ab initio systems, showing promise for applications to periodic systems using, e.g.\ GW \cite{Hedin:1965aa, 10.3389/fchem.2019.00377}.
The compact real-time representation may also find applications in quantum computing, where the required number of measured observables scales with the number of time points \cite{Ortiz:2001aa, Wecker:2015aa, Bauer:2016aa}.

Finally, the success of the real-time panel expansion, shown here for \emph{equilibrium} real-time evolution, is an important first step towards high-order expansion methods for \emph{non-equilibrium} real-time evolution.
The presented discretization of the mixed Green's function $G^\ttau(t,\tau)$ is directly applicable to the non-equilibrium case, while the generalization of the high-order expansion idea to the two real-time dependent Green's function components, e.g.\ $G^\lessgtr(t, t')$ [Eq.\ (\ref{eq:G_gtr_contour_component}, \ref{eq:G_les_contour_component})], is yet to be explored.

% --------------------------------------------------------------------
% --------------------------------------------------------------------
\begin{acknowledgments}
% --------------------------------------------------------------------
%
The authors would like to acknowledge helpful discussions with D.\ Zgid on quantum chemistry applications, J.\ Kaye on the numerical intricacies of orthogonal polynomial expansions, M.\ Gulliksson on superconvergence, and S.\ Iskakov on EDLib \cite{Iskakov:2018aa}. We thank P.~Pavlyukh for spotting and helping us correct two sign typos.
%
%Part of this material is based upon work supported by the U.S. Department of Energy, Office of Science, Office of Advanced Scientific Computing Research and Office of Basic Energy Sciences, Scientific Discovery through Advanced Computing (SciDAC) program under Award Number DE-SC0022088.
%
The work of E.G.\ was supported by the U.S.\ Department of Energy, Office of Science, Office of Advanced Scientific Computing Research and Office of Basic Energy Sciences, Scientific Discovery through Advanced Computing (SciDAC) program under Award Number DE-SC0022088. 
H.U.R.S.\ acknowledges funding from the European Research Council (ERC) under the European Union’s Horizon 2020 research and innovation programme (Grant agreement No.\ 854843-FASTCORR).
The computations were enabled by resources provided by the Swedish National Infrastructure for Computing (SNIC) through the projects SNIC 2020/5-698 and SNIC 2020/6-294 at the High Performance Computing Center North (HPC2N) partially funded by the Swedish Research Council through grant agreement no.\ 2018-05973.
\end{acknowledgments}

% --------------------------------------------------------------------
\appendix
% --------------------------------------------------------------------

% --------------------------------------------------------------------
\section{Imaginary time Volterra integral} \label{app:imtime}
% --------------------------------------------------------------------

The convolution operator in Eq.\ (\ref{eq:mat_B}) derived in Ref.\ \onlinecite{doi:10.1063/5.0003145} pertains to the imaginary time convolution integral
\begin{align}
  [A {\ast}] B \equiv
  \int_0^\beta d\bar{\tau} A(\tau - \bar{\tau}) B(\bar{\tau})
  \, .
  \label{eq:imtimecov}
\end{align}
Comparing with the imaginary-time integral for the right-hand side term $Q^\ttau_p(t, \tau)$ in Eq.~\ref{eq:Qp} we have
\begin{equation}
  A(\tau) \equiv
  G^M(-\tau)
  \, , \quad
  B(\tau) \equiv \Sigma^{\ttau}_p(t, \tau)
  \, .
\end{equation}
The fermionic antiperiodicity $G^M(-\tau) = -G^M(\beta - \tau)$ in combination with the Legendre expansion of $G^M(\tau)$ in Eq.\ (\ref{eq:taubasis}) gives
\begin{align}
  A(\tau)
  &= -\sum_m G^M_m P_m[\psi_M(\beta - \tau)]
  \nonumber \\
  &= \sum_m (-1)^{m+1} G^M_m P_m[\psi_M(\tau)]
  \nonumber \\
  &= \sum_m A_m P_m[\psi_M(\tau)]
  \, ,
  \label{eq:A_leg}
\end{align}
where we have used that $\psi_M(\beta - \tau) = - \psi_M(\tau)$, see Eq.\ (\ref{eq:tau_psi}), and $P_m(-x) = (-1)^m P_m(x)$.
Hence, with the imaginary time convolution operator in Eq.\ (\ref{eq:mat_B}), the panel Legendre expansion of $Q^\ttau_p(t, \tau)$ can be expressed as
\begin{equation}
  Q^{\ttau}_{p, nm}
  =
  \sum_{m'} [A{\ast}]_{mm'} \Sigma^\ttau_{p, nm'}
  \, .
\end{equation}
where the convolution operator $[A {\ast}]$ is built using the Legendre coefficients $A_m$ of $A(\tau)$ given by $A_m = (-1)^{m+1} G^M_m$ in Eq.\ (\ref{eq:A_leg}).

% --------------------------------------------------------------------
\section{Legendre polynomial sparse sampling in Matsubara frequency} \label{app:SparseSampling}
% --------------------------------------------------------------------

% --------------------------------------------------------------------
% Figure: Sparse sampling grid
% --------------------------------------------------------------------
\begin{figure}[tbh]
\includegraphics[scale=1] {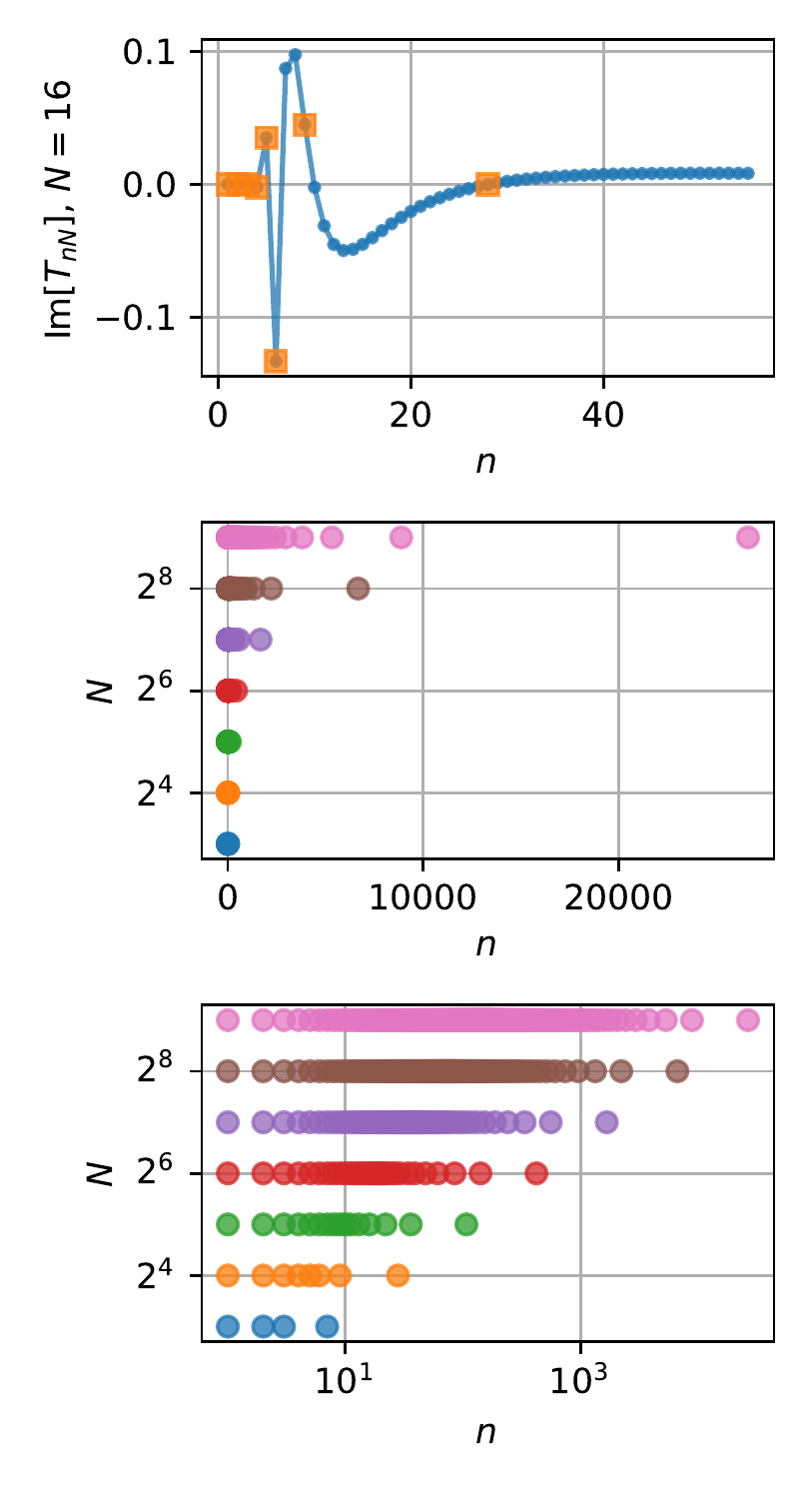}
\caption{\label{fig:sparse} Sparse-sampling Matsubara frequency grids based on Legendre polynomials of order $N$. Upper panel: Matsubara frequency transform of the Legendre polynomial $P_N(x)$ for $N=16$ (blue dots) and the sparse-sampling frequencies (orange squares). The selected Matsubara frequency indices $n$ are shown for different orders $N$ on a linear (middle panel) and logarithmic grid (lower panel).
} \end{figure}
% --------------------------------------------------------------------

The $N^\text{th}$ order Legendre-Gauss quadrature nodes $x_i$ can be constructed as the roots of the $N^\text{th}$ Legendre polynomial, $P_N(x_i) = 0$. Sparse sampling in Matsubara frequency takes this idea to the imaginary frequency axis. The approach has previously been applied to Chebyshev polynomials \cite{PhysRevB.101.035144} and here we extend the approach to Legendre polynomials.

The Fourier transform of Legendre polynomials
\begin{equation}
  \int_{-1}^1 e^{ia x} P_l(x) = 2 i^l j_l(a)
%  \, ,
\end{equation}
can be used to construct the linear transformation \cite{Boehnke:2011fk} $T_{nl}$ from Legendre coefficients to Matsubara frequencies $\omega_n$,
\begin{equation}
  T_{nl} = (-1)^n i^{l + \zeta} j_l\left( \frac{\pi(2n + \zeta)}{2} \right)
  \, .
\end{equation}
The Matsubara frequency sampling points can therefore be selected as the $N$ first points where the linear transform $T_{nN}$ of the $N^\text{th}$ order Legendre polynomial changes sign.

The resulting Matsubara frequency grids selects a number of equidistant Matsubara frequencies at low frequencies and only a few (non-linearly spaced) points at high-frequency, see Fig.\ \ref{fig:sparse}.
%\ \\

% --------------------------------------------------------------------
\section{Interaction energy and spectra for He$_2$} \label{app:He2}
% --------------------------------------------------------------------

% --------------------------------------------------------------------
% Figure: He2 Interaction Energy
% --------------------------------------------------------------------
\begin{figure}
\includegraphics[scale=1]{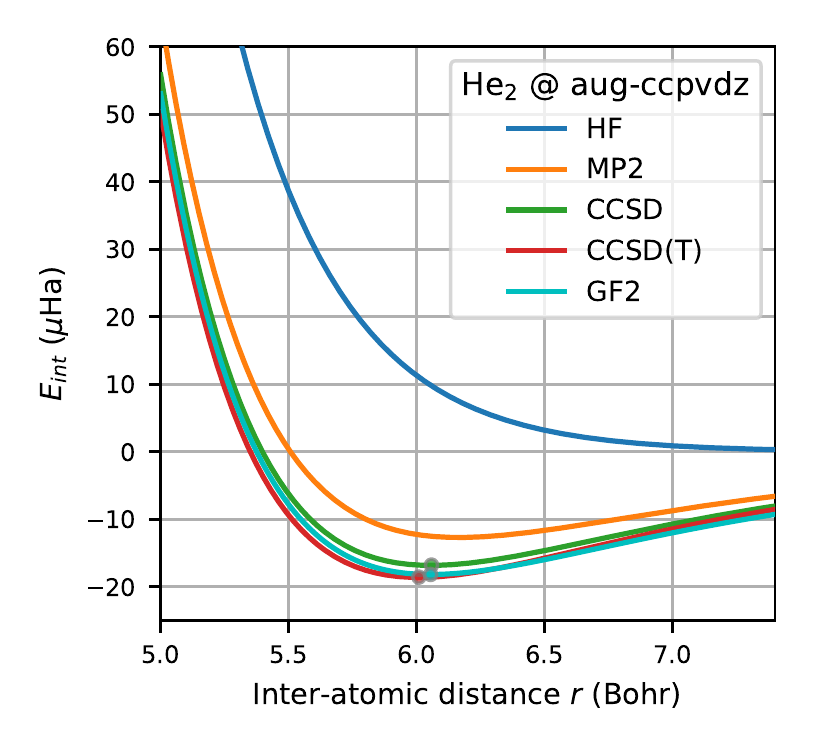}\\[-5mm]
\caption{\label{fig:He2_Eint} Interaction energy $E_{\text{int}} \equiv E_{\text{He}_2} - 2 E_{\text{He}}$ (counterpoise corrected \cite{doi:10.1080/00268977000101561}) for He$_2$ using the aug-ccpvdz basis (solid lines), as a function of interatomic distance $r$ for HF, MP2, CCSD, CCSD(T) and GF2 (at $\beta=200\,\text{Ha}^{-1}$), the minima of CCSD and CCSD(T) are from Ref.\ \cite{Van-Mourik:1999aa}) and the GF2 minima from Ref.\ \cite{doi:10.1063/5.0003145} (markers).
} \end{figure}
% --------------------------------------------------------------------

% --------------------------------------------------------------------
% Figure: He2 Spectral function
% --------------------------------------------------------------------
\begin{figure*}
\includegraphics[scale=1]{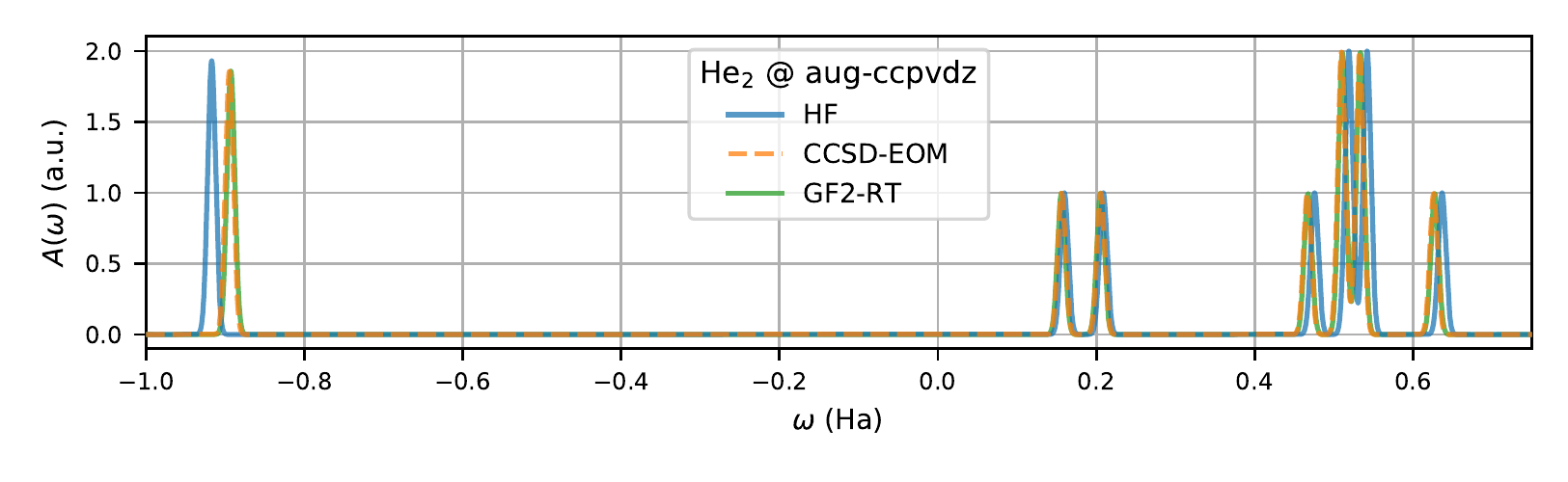}\\[-5mm]
\caption{\label{fig:He2_Spectra} Spectral function of He$_2$ using the aug-ccpvdz basis from HF, CCSD-EOM, and GF2-RT.
} \end{figure*}
% --------------------------------------------------------------------

The performance of GF2 in the covalently bound systems $H_2$ and $LiH$ reported in the main text are very different compared to the case of the noble gases. As an example we reproduce the result on the diatomic interaction energy of He$_2$ from Ref.\ \onlinecite{doi:10.1063/5.0003145} in Fig.\ \ref{fig:He2_Eint}. For He$_2$ the interaction energy of GF2 constitutes a drastic improvement compared to MP2, lying in between the CCSD and CCSD(T) results in a region around the equilibrium atomic separation.

With the equilibrium real-time propagation we can now compare the spectral functions for He$_2$ from HF, CCSD-EOM and GF2-RT, see Fig.\ \ref{fig:He2_Spectra}. The GF2-RT result agrees quantitatively with CCSD-EOM while HF gives dissernable shifts and an amplitude change in the occupied resonance at $\omega \approx -0.9\,\text{Ha}$, comprised of two near degenerate molecular orbitals.

% --------------------------------------------------------------------
\section{Extended Koopman's Theorem (EKT)} \label{app:EKT}
% --------------------------------------------------------------------

Koopman's theorem \cite{Koopmans:1934wi} can -- in Hartree-Fock quadratic mean-field theory -- be used to approximate single-particle excitation energies like the ionization potential (IP) and electron affinity (EA) by the single-particle eigenstates of the mean field Hamiltonian.

The extension to higher order correlated methods is called the extended Koopman's theorem (EKT) \cite{Smith:1975ww, Day:1975wm, Morrison:1975ux, Ellenbogen:1977uh, Chipman:1977va, Vanfleteren:2009ul}. EKT is based on the generalized Hartree-Fock one-particle potentials $V^\lessgtr$ and their corresponding generalized overlap matrices $S^\lessgtr$, where lesser $<$ and greater $>$ denotes the occupied and unoccupied states, respectively.

In Green's function based methods the matrices $V^\lessgtr$ and $S^\lessgtr$ are determined by the imaginary-time Green's function $G(\tau)$ according to \cite{Dahlen:2005aa, Stan:2006vc, PhysRevB.97.115164}
\begin{equation}
  S^\lessgtr = - G(\tau) \Big|_{\tau = 0^\pm}
  \, , \quad
  V^\lessgtr = \partial_\tau G(\tau) \Big|_{\tau = 0^\pm}
  \, .
\end{equation}
The eigenstates $\psi^\lessgtr_a$ of the related generalized eigenvalue problem
\begin{equation}
  V^\lessgtr \cdot \psi^\lessgtr_a = \epsilon^\lessgtr_a \, S^\lessgtr \cdot \psi^\lessgtr_a
  \, ,
\end{equation}
are the variationally stable natural transition orbitals with eigen-energies $\epsilon^\lessgtr_a$.

Using the natural transition orbitals, the ionization potential $E_{IP}$ and electron affinity $E_{EA}$ can be approximated as
\begin{equation}
  E_{IP} = - \max \epsilon^<_a
  \, , \quad
  E_{EA} = - \min \epsilon^>_a
  \, ,
\end{equation}
and the occupied and unoccupied single particle spectral functions $A^\lessgtr$ can be approximated as
\begin{equation}
  A^\lessgtr(\omega) \approx
  \sum_a | S^\lessgtr \cdot \psi^\lessgtr_a |^2 \delta(\omega - \epsilon^\lessgtr_a)
  \, ,
\end{equation}
which gives the total single-particle spectral function $A(\omega)$ as $A(\omega) = A^>(\omega) + A^<(\omega)$.

% --------------------------------------------------------------------
\bibliography{manuscript}
\end{document}